\documentclass{sig-alternate-05-2015}

\usepackage{amsmath}
\usepackage{dsfont}
\usepackage{scrextend}
\usepackage{balance}  
\usepackage{verbatim}
\usepackage{amssymb}
\usepackage{mathrsfs}
\usepackage{hyperref}
\usepackage[usenames,dvipsnames]{color}
\usepackage{multirow}
\usepackage{graphicx}
\usepackage{caption}
\usepackage{subcaption}
\usepackage{enumitem}
\usepackage{booktabs} 
\usepackage{url}
\usepackage{xspace}
\usepackage{outlines}
\usepackage{wasysym}
\usepackage{adjustbox}
\usepackage[ruled, vlined, linesnumbered]{algorithm2e}

\usepackage{etoolbox} 

\newcommand{\TableCorpus}{\mathcal{T}}
\newcommand{\TwoColTblCorpus}{\mathcal{B}}
\newcommand{\PartitionSet}{\mathcal{P}}

\newcommand{\argmax}{arg\,max}
\newcommand{\web}{\texttt{Web}}
\newcommand{\ent}{\texttt{Enterprise}}

\newtoggle{fullversion}
\toggletrue{fullversion}

\newtheorem{thm}{Theorem}

\newtheorem{ndef}[thm]{Definition}

\newtheorem{nexp}[thm]{Example}
\newtheorem{example}[thm]{Example}
\newtheorem{problem}[thm]{Problem}

\begin{document}

\title{Synthesizing Mapping Relationships Using Table Corpus}

\numberofauthors{2}
\author{
\alignauthor Yue Wang\titlenote{Work done at Microsoft Research.} \\
       \affaddr{University of Massachusetts Amherst}\\
       \affaddr{Amherst, MA, USA}\\
       \email{yuewang@cs.umass.edu}
\alignauthor Yeye He \\
       \affaddr{Microsoft Research}\\
       \affaddr{Redmond, WA, USA}\\
       \email{yeyehe@microsoft.com}
}

\maketitle

\begin{abstract}
Mapping relationships, such as (\texttt{country}, 
\texttt{country-code}) or (\texttt{company},
\texttt{stock-ticker}), are versatile data assets 
for an array of applications in 
data cleaning and data integration 
like auto-correction and auto-join. However, today 
there are no good repositories of mapping tables that can
enable these intelligent applications.

Given a corpus of tables such as web tables or
spreadsheet tables, we observe that values of
these mappings often exist in pairs of columns in same tables.
Motivated by their broad applicability, we study the problem of synthesizing
mapping relationships using a large table corpus.
Our synthesis process leverages 
compatibility of tables based on co-occurrence statistics, 
as well as constraints such as functional dependency. 
Experiment results using web tables and enterprise spreadsheets 
suggest that the proposed approach can produce high quality 
mappings.

\end{abstract}


\section{Introduction} \label{sec:intro}
\textit{Mapping tables}, sometimes also referred to as 
\textit{bridge tables}~\cite{Kimball02},
are two-column tables where each distinct value in 
the left column maps to a unique value 
in the right column (or
functional dependencies hold).
Table~\ref{tab:examples} gives a few example mapping tables
with one-to-one mapping relationships.
Table~\ref{tab:examples2} shows additional examples
with many-to-one mappings.


\begin{table}
    \centering
	\hspace{-11mm}
	\begin{subfigure}[b]{0.23\textwidth}
    		\centering
		\includegraphics[height=0.85in]{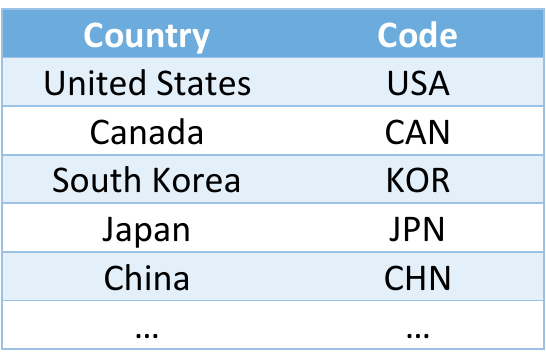}
		\caption{ISO country codes} \label{tab:country}
	\end{subfigure}
	\hspace{-3mm}
	\begin{subfigure}[b]{0.23\textwidth}
		\centering
		\includegraphics[height=0.85in]{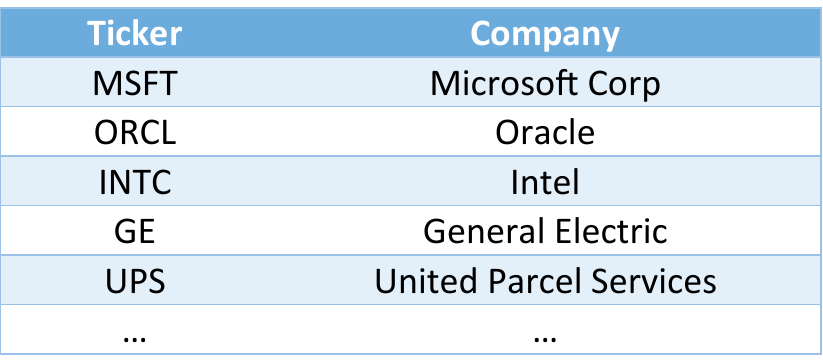}
		\caption{Stock tickers} \label{tab:ticker}
	\end{subfigure}\\
	\hspace{-11mm}
	\begin{subfigure}[b]{0.23\textwidth}
    		\centering
		\includegraphics[height=0.85in]{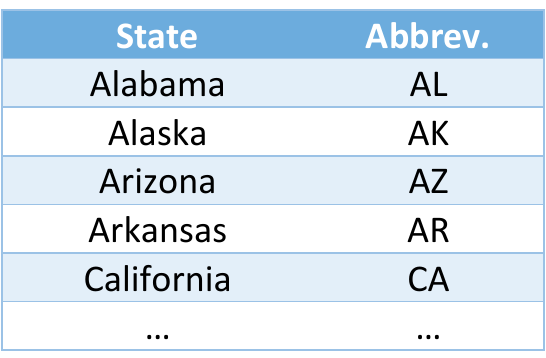}
		\caption{State abbreviations}\label{tab:state}
	\end{subfigure}
	\hspace{-3mm}
	\begin{subfigure}[b]{0.23\textwidth}
		\centering
		\includegraphics[height=0.85in]{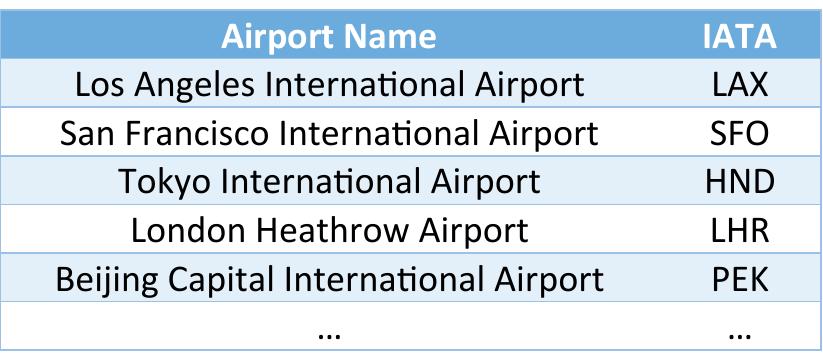}
		\caption{Airport IATA codes}\label{tab:airport}
	\end{subfigure}
\vspace{-0.2mm}
\caption{Example one-to-one mapping tables: (a) Countries to ISO codes, 
(b) Company names to stock-tickers, (c) State names to abbreviations,
(d) Airports to IATA-codes.}\label{tab:examples}
\end{table}

\begin{table}
    \centering
	\begin{subfigure}[b]{0.23\textwidth}
    		\centering
		\includegraphics[height=0.85in]{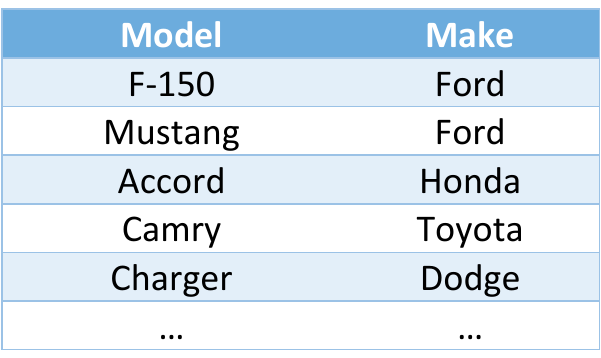}
		\caption{Car make and model}
	\end{subfigure}
	\begin{subfigure}[b]{0.23\textwidth}
		\centering
		\includegraphics[height=0.85in]{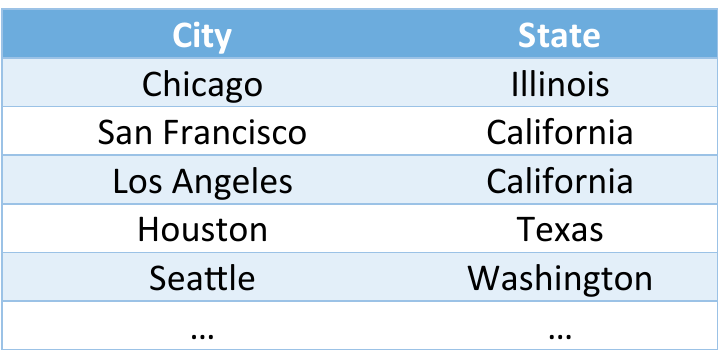}
		\caption{City and state} \label{tab:city}
	\end{subfigure}
\vspace{-2.0ex}
\caption{Example many-to-one mapping tables: (a) Car makes and models, 
(b) Cities and states}\label{tab:examples2}
\end{table}

\begin{table}
  \centering
  \begin{minipage}[b]{0.23\textwidth}
    \includegraphics[height=0.83in]{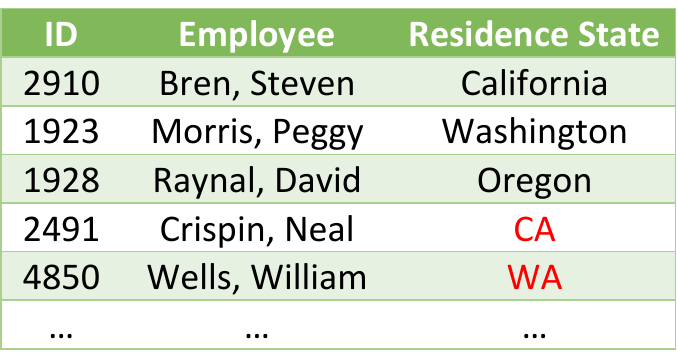}
    \vspace{-1.0ex}
    \caption{Auto-correction: correct inconsistent values 
(highlighted) using Table~\ref{tab:state}.} 
	\label{tab:auto-correction}
  \end{minipage}
  \hfill
  \begin{minipage}[b]{0.23\textwidth}
    \includegraphics[height=0.83in]{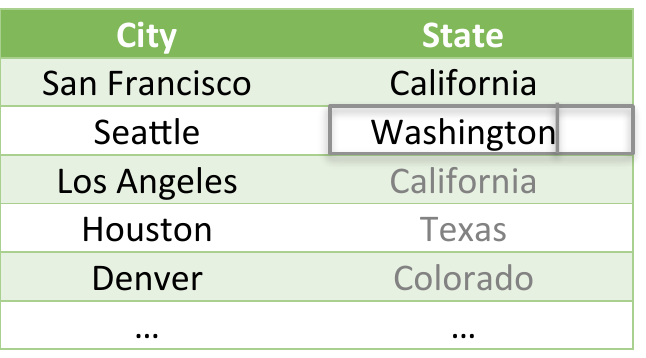}
    \vspace{-1.0ex}
    \caption{Auto-fill: automatically populate values 
based on mappings from Table~\ref{tab:city}.} 
	\label{tab:auto-fill}
  \end{minipage}
\end{table}

\begin{table}
		\centering
		\includegraphics[height=0.85in]{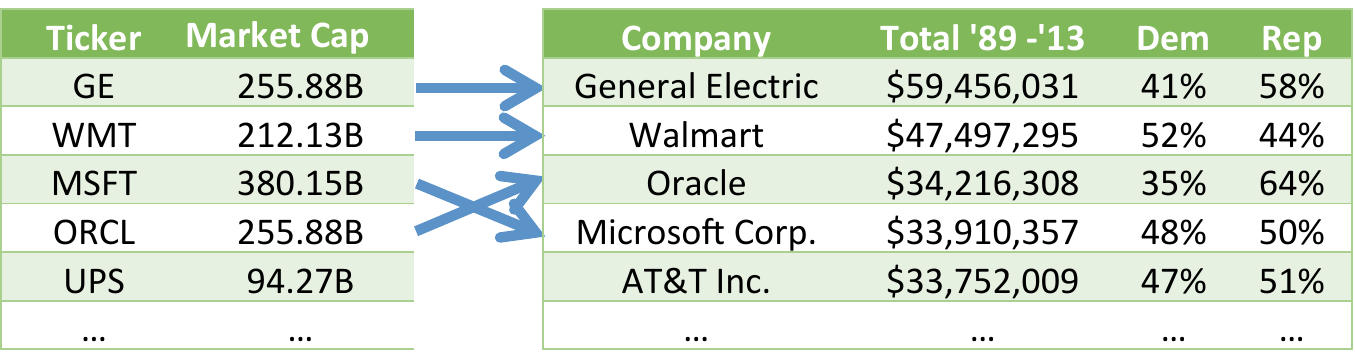}
    \vspace{-1.0ex}
		\caption{Auto-join: joining related tuples based on mappings
from Table~\ref{tab:ticker}}
		\label{tab:auto-join}
\end{table}

Mapping tables like these are important data 
assets for a variety of applications 
such as data integration and data cleaning. We briefly discuss three
scenarios here.

\texttt{Auto-correction}. Real-world tables are often dirty, where 
inconsistent values may be present in same columns. 
Table~\ref{tab:auto-correction} shows such an example. The last
column about \texttt{state} are mixed with both full state names 
and state abbreviations. An 
intelligent data quality agent, equipped with the mapping table 
in Table~\ref{tab:state}, can easily detect and alert users about
such inconsistency, by discovering that values in the left and right column of
Table~\ref{tab:state} are mixed in one user data column. Furthermore, it can
automatically suggest corrections based on the mapping
relationship (e.g., correcting \texttt{CA} to \texttt{California}).

\texttt{Auto-fill}. In this example scenario in 
Table~\ref{tab:auto-fill}, a user has a list of city
names. She wants to add a column of state names
corresponding to the cities. By just entering
a few example values (e.g., \texttt{California} for 
\texttt{San Francisco}), the system automatically 
discovers the intent by matching existing value pairs 
with those in Table~\ref{tab:city}, and can thus suggest to automatically
fill remaining values in the right column (grayed out 
in Table~\ref{tab:auto-fill}).

\texttt{Auto-join}. In data integration and ad-hoc data analysis, users often need
to ``join'' two tables together, whose key columns may have different
representations. In Table~\ref{tab:auto-join} for example, an analyst needs to join 
the left table that has stocks by their market capitalization, with the right table that lists
companies by their political contributions, to analyze potential correlations. 
However a direct join is not possible since the subject column of 
the left table is stock tickers, while the right table 
uses company names. A system equipped with mapping
tables would make the join possible by using
Table~\ref{tab:ticker} as an intermediate bridge that performs 
a three-way join to connect these two user tables,
without asking users to provide an explicit mappings.

\textbf{Synthesize mapping tables with human curation.} 
In this work, we develop methods to automatically
synthesize mapping relationships from existing table corpora, where the
goal is to generate as many high-quality mappings as possible.
Because algorithms are 
bound to make mistakes, additional human verification and curation
can be used to ensure very high precision (Section~\ref{sec:curation}). 
The resulting mappings can then be
utilized to enable the applications discussed above 
in a unified manner.

\textbf{Why pre-compute mappings.}
While there are separate solutions for auto-join and
auto-fill problems (e.g.,~\cite{He15, Yakout12}),
our approach has a few important advantages. 

First, synthesized mappings are 
amenable to human inspection and curation, which is 
critical to ensure
very high quality. In attempting to commercialize
technologies similar to~\cite{He15, Yakout12} in 
enterprise spreadsheet software like Excel, 
the main feedback we received is the trustworthiness 
of results produced by black-box algorithms. Algorithms with even
$99\%$ correctness is still unacceptable in the
context of enterprise spreadsheets, because any error
introduced by algorithms would be difficult for users to detect,
but is highly embarrassing and damaging in enterprise settings. 

An analogy we would like to draw is the knowledge-bases used in search engines
such as Google and Microsoft Bing. 
Similar to our problem, the quality required for knowledge-bases
is also very high, so commercial
knowledge bases are created in offline processes
that combine algorithmic automation 
with human curation. Mapping tables 
can be viewed as the counterpart of knowledge bases in the relational world,
where a similar curation process may be needed because of the quality requirement.
And like search engines that have millions of users, 
spreadsheet software can reach millions of data analysts, such that the 
cost of curating mappings can be amortized over a large user base to make the 
effort worthwhile.

Second, synthesized mapping relationships can be materialized as tables, which are
easy to index and efficient to scale to large problems.
For example, instead of performing expensive online
reasoning over large table corpora for specific applications like auto-join~\cite{He15} and 
auto-fill~\cite{Yakout12},
one could index synthesized mapping tables using hash-based techniques 
(e.g., bloom filters) for efficiently lookup based on value containment. 
Such logic is both simple to implement and easy to scale.

Lastly, mapping tables are versatile data assets with many applications. 
By solving this common underlying problem
and producing mapping tables as something that can be easily plugged
into other applications, 
it brings benefits to a broad class of applications as opposed to requiring 
separate reasoning logic to be developed for different applications (e.g., \cite{He15} for auto-join and \cite{Yakout12}
for auto-fill).

\textbf{Why synthesize tables.}
Given table corpora such as HTML tables from web or
spreadsheets from enterprises, fragments of useful mapping relationships 
exist.  For example, the \texttt{country} and \texttt{country-ISO3-code}
columns in Table~\ref{tab:country} are often 
adjacent columns in same tables on the web.
As such, an alternative class of approaches is to
``search'' tables based on 
input values and then ask users to select relevant
ones (e.g., Google Web Tables~\cite{googlewebtables}, 
Microsoft Power Query~\cite{powerbi}, and DataXFormer~\cite{Abedjan:2015}). 
However, because desired values pairs
often span across multiple tables,
users frequently need to search, inspect and 
understand table results, before manually
piecing them together from multiple tables. Our experience suggests 
that this process is often too cumbersome for end users.

Mappings synthesized from multiple tables, on the other hand, take 
away the complexity and make it 
easy for end users. More specifically, synthesized mappings
have the following benefits.

\noindent $\bullet$  \textit{Completeness}. In many
cases one table only covers a small fraction of mappings in the
same relationship. For example, while there exist
thousands of airports, a web table like Table~\ref{tab:airport}
often lists only a small fraction of popular airports. Stitching together 
tables in the same relationship provides better coverage and is clearly desirable. 

\noindent $\bullet$  \textit{Synonymous mentions}. Each individual table 
from a table corpus typically only has one
mention for the same entity. For example, Table~\ref{tab:country} has
\texttt{South Korea} and \texttt{KOR}. 
In reality different tables use different 
but synonymous names. Table~\ref{tab:synonym}
shows real results synthesized from many web tables,
which has different synonyms of \texttt{South Korea}.
Similarly the right part has many synonyms 
for \texttt{Congo}. Note that a specific synonym of \texttt{South Korea}
may not necessarily co-occur with another synonym of \texttt{Congo}
in the same web table, and the probability of co-occurrence in conjunction 
with synonyms of additional countries is even lower. 
However, any combination of these synonyms may 
actually be used in user tables that may require auto-join or auto-fill.
Using single tables as mappings would not provide sufficient coverage in these cases.
On the other hand, if all these synonyms are synthesized together
as one table like in Table~\ref{tab:synonym}, then any combination 
of these synonyms can still be covered without requiring users to perform manual
synthesis from multiple tables.

\noindent $\bullet$  \textit{Spurious mappings}. Certain mappings that
appear to hold locally in single tables may not be meaningful. 
For example, a random table listing \texttt{departure-airport}
and \texttt{arrival-airport} may happen to have values observe 
functional dependency at the instance level.
However, at a conceptual level this is not a useful mapping. 
Such a spurious mapping, when indexed from
single tables, can trigger false-positive results
for applications like auto-correct and auto-join.
A holistic analysis of global relationships are necessary 
to identify true mappings from spurious ones.

\textbf{Existing approaches:} Given that table synthesis
is needed to assist human curation, we look at existing techniques that can be used here.


\begin{table}
		\centering
		\includegraphics[height=1.06in]{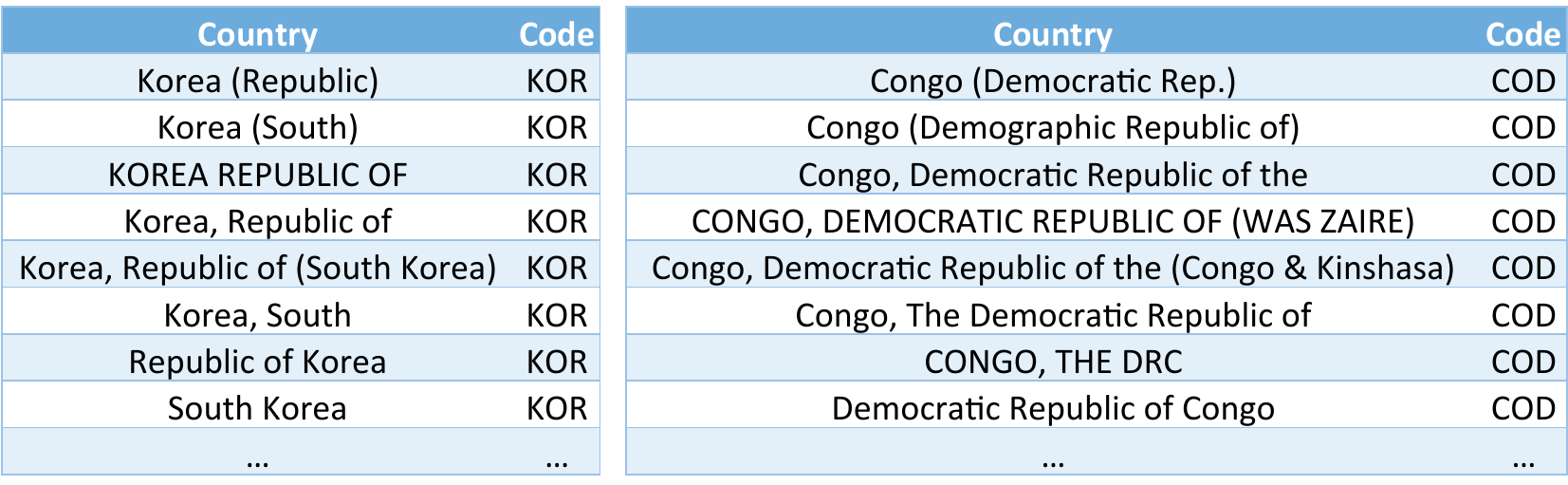}
		\caption{Examples from a synthesized mapping relationship (\texttt{country}, \texttt{country-ISO3-code}) using real web tables. 
The left table shows examples of synonyms for the country \texttt{South Korea}, all
of which map to the same code \texttt{KOR}. The right table shows similar examples for 
\texttt{Congo}.}
	\label{tab:synonym}
\end{table}

\texttt{Union tables.} Ling and Halevy et al. studied the problem of
stitching together web tables in the same web-domain 
(where tables are more homogeneous)
based on meta data such as column 
names~\cite{Ling:2013}. While the technique is 
not designed to synthesize relationships from a large heterogeneous corpus, 
it is the only work we are aware of that 
performs table synthesis from corpora.
We will show that adapting this to a large corpus of heterogeneous tables
will fail, because column names are often undescriptive~\cite{Cortez15}
that leads to over-grouping and low-quality mappings.
For example, in Table~\ref{tab:country}, the
column name for countries are often just \texttt{name}, 
and the column name for country-codes may 
be \texttt{code}. As a result, grouping by column names
tends to lump this table with other name-to-code mappings.
Our approach reasons about compatibility of tables based on values,
which are more reliable in telling the true relationships.

\texttt{Schema matching.} 
There is a long and fruitful line of research on schema 
matching that suggests possible 
mappings between table columns~\cite{Rahm01}.
However, schema matching is typically used in database contexts
for a small number of schemas, and produces pair-wise
matches  for human users to evaluate. In our problem
we are given hundreds of millions of schemas as input,
for which pairwise human verification is infeasible, and
aggregation of pairwise decisions to a group level 
is necessary for human curation. Furthermore, since
we are only interested in mapping relationships,
which are a specific type of tables that always observe functional dependencies,
we can derive additional \textit{negative incompatibility} 
induced by FDs that is not explored by schema matching.  
For example, there are multiple country-to-code relationships 
such as (\texttt{country} $\rightarrow$ \texttt{ISO3-country-code}),
 (\texttt{country} $\rightarrow$ \texttt{FIFA-country-code}), 
 (\texttt{country} $\rightarrow$ \texttt{IOC-country-code}), etc,
all of which share substantial value overlap as well 
as similar column names. Schema matching techniques
would identify them as matches and merge them 
incorrectly, whereas we would prevent the synthesis because 
of the FD-based incompatibility.  Considering both positive
and negative signals is critical
for high-quality synthesis at a large scale.

\texttt{Knowledge base.} Knowledge bases (KB) such 
as Freebase~\cite{Bollacker:2008} and YAGO~\cite{Suchanek:2007}
have important entity-relationships that
can be viewed as synthesized (semi-automatically) from different
sources. However, many mappings are missing from KB.
For instance, YAGO has none of the example mappings listed
in Table~\ref{tab:examples} (all of which are common mappings), 
while Freebase misses two (stocks and airports). Furthermore, 
for mappings that do exist in KB, they typically do not have
synonyms like the ones in Table~\ref{tab:synonym}.
Lastly, KB have limited coverage beyond the public web domain,
such as mapping (\texttt{cost-center-name} $\rightarrow$
\texttt{cost-center-code}) that is specific to enterprises domains.

\textbf{Contribution.} Observing that mapping relationships
are well-represented in tables, we propose to automatically 
synthesize mapping relationships using table 
corpora. We formalize this as an optimization problem that maximizes 
positive compatibility between tables while respecting
constrains of negative compatibility
imposed by functional dependencies. We show a trichotomy of
complexity for the resulting optimization problem, and develop an efficient
algorithm that can scale to large table corpus 
(e.g., 100M tables). 
Our evaluation using real table corpora suggests that the proposed 
approach can synthesize high quality mapping tables.


\section{Solution Overview} \label{sec:framework}

In this section, we first introduce notions like mapping relationships 
and table corpora necessary for discussions. We then give a high-level overview of our  synthesis solution.

\begin{figure}
\centering
\includegraphics[width=1.\columnwidth]{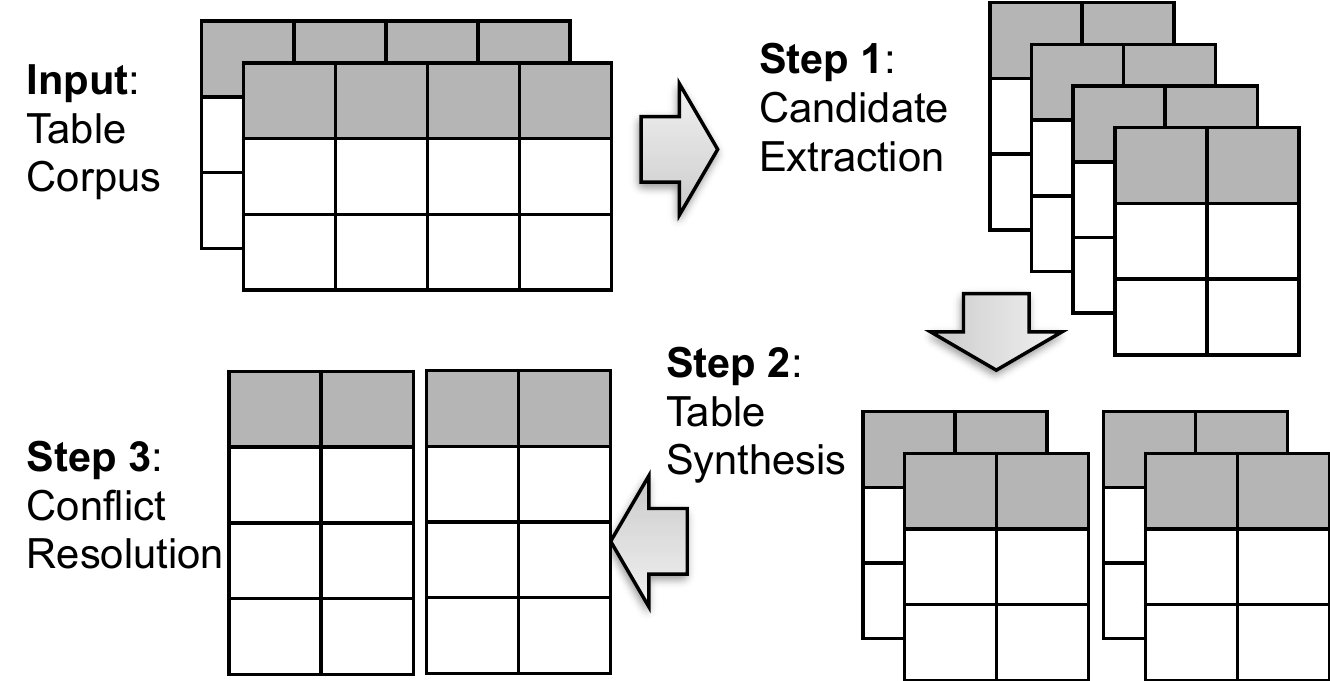}
\caption{Solution overview with three main steps: (1) Extract
candidate two-column-tables; (2) Synthesize related tables; (3) Resolve
conflicts in the same relationship.}\label{fig:framework}
\vspace{-2mm}
\end{figure}

\subsection{Preliminaries}

\textbf{Mapping relationships}. 
The goal of this work is to discover mapping relationships. 
Specifically, we focus on binary mappings involving two attributes.

\begin{ndef}
Let $R$ be a conceptual relation with two attributes $X, Y$.
The relationship is a \textit{mapping relationship}, 
denoted by $M(X, Y)$ or $X \rightarrow Y$, 
if for all $x \in X$, x functionally determines
one and precisely one value $y \in Y$.
\end{ndef}

Examples of mapping relationships 
include (\texttt{country} $\rightarrow$
\texttt{country-code}) and (\texttt{company} $\rightarrow$ \texttt{stock-ticker}) 
as shown in Table~\ref{tab:examples} and Table~\ref{tab:examples2}. 
There is a mapping relationship between attributes
\texttt{country} and \texttt{country-code}, for instance,
since value in one attribute is uniquely associated with precisely one value
in the other attribute.

Note that this is closely related to 
functional dependency (FD), traditionally defined over one physical table.
We make the distinction to define mappings as 
conceptual relationships that can be represented
in multiple tables, but may never be fully embodied in one physical table (e.g.,
the synthesized mapping shown in Table~\ref{tab:synonym} 
with both \texttt{South Korea} and \texttt{Korea (South)}
would not occur in one table).

Existing FD discovery work mainly focuses on efficiency (e.g.,~\cite{Huhtala99}),
because it is intended for interactive data exploration on ad-hoc 
data sets. However, in our problem the key challenge is to produce high-quality
synthesis of tables to assist human curation, where efficiency is not as
important because the corpus is given a priori and synthesis can be run as offline jobs.

For cases where both $X \rightarrow Y$ and $Y \rightarrow X$
are mapping relationships, we call such bi-directional relationships
1:1 mappings (examples are in Table~\ref{tab:examples}). If the mapping
relationship only holds in one direction, then it is an N:1 mapping
(Table~\ref{tab:examples2}).

It is worth noting that in practice, because of name ambiguity,
functional relationship in some mappings may appear
to only hold approximately. For example,
\texttt{city} $\rightarrow$ \texttt{state} is conceptually
a mapping relationship. However, when entities are represented 
as strings, the functional relationship may not completely hold.
For example, in the same table there may be a city called \texttt{Portland} in 
the state of \texttt{Oregon}, and another city \texttt{Portland} in 
the state of \texttt{Maine}, thus giving the appearance of violating
FD. To take such name
ambiguity into account, we consider relationships whose surface forms 
are approximate mapping relationships.

\begin{ndef}
Let $R$ be a conceptual relation with two attributes $X, Y$.
The relationship is a \textit{$\theta$-approximate 
mapping relationship}, 
denoted by $M_{\theta}(X, Y)$ or $X \rightarrow_{\theta} Y$, 
if there exists a subset $\overline{R} \subset R$ with 
$|\overline{R}| \geq \theta |R|$, in which all $x \in X$ 
functionally determines one and precisely one value $y \in Y$.
\end{ndef}

We consider approximate mappings with $\theta$ over 95\%.
Hereafter we will simply use mapping
relationship to refer to its $\theta$-approximate
version when the context is clear.

\textbf{Table corpora.} 
The only input to our problem is a corpus of tables.
\begin{ndef}
\label{def:approx}
A table corpus $\TableCorpus = \{T\}$ is a set of 
relational tables $T$, each of which consists of a set of columns,
or written as $T = \{C_1, C_2, ... \}$.
\end{ndef}

Today relational tables are abundant and are very 
rich in nature. In this study, we use a corpus of 100M tables extracted
from the Web, and a corpus of 500K tables extracted
from spreadsheet files crawled from the intranet of 
a large enterprise.

\subsection{Solution Overview}

Our approach has three main steps, as shown in Figure~\ref{fig:framework}.

\vspace{-2mm}
\paragraph*{Step 1: Candidate Extraction} This step starts by exhaustively 
extracting pairs of columns from all tables in the corpus as candidates 
for synthesis. For each table $T = \{C_1, C_2, ..., C_n \}$ with $n$ columns, 
we can extract $2{{n}\choose{2}}$ such ordered pairs. 
However, many column pairs are not good candidate for 
mapping relationships because (1) for some column pair 
if the local relationship is already not functional,
then it is unlikely to participate in true mappings;
and (2) some table columns are of low quality and are not coherent enough
(e.g., with mixed concepts). To address these issues, we use
FD constraints as well as value-based co-occurrence
statistics to prune away low-quality candidate tables.

\vspace{-2mm}
\paragraph*{Step 2: Table Synthesis} In this step, we
judiciously synthesize two-column 
tables that describe the same relationship and are compatible with
each other. The reason this is necessary is because many web tables
and spreadsheets are for human consumption~\cite{Ling:2013}, 
and as a result contain only a subset of instances for the ease of browsing.
Furthermore, one table in most cases mentions an entity by
one name; synthesis helps to improve coverage of synonyms that are
important for many applications.

\vspace{-2mm}
\paragraph*{Step 3: Conflict Resolution} Because results from table 
synthesis piece together many tables, some of which are bound to
have erroneous values inconsistent with others, namely two pairs
of values in the same mapping with the same left-hand-side value
but different right-hand-side (thus violating the definition of mappings).
These can often happen due to quality issues
or extraction errors. We apply a post-processing step 
to resolve conflicts in synthesized mapping relationships to produce
our final results.


\section{Candidate Table Extraction}
\label{sec:preprocessing}

In this section we briefly describe the preprocessing of tables. 
Recall that in this work we focus on synthesizing binary
mapping relationships. We start with
two-column tables extracted from an existing table corpus.
Given a table $T = \{C_1, C_2, ... C_n \}$ with $n$ columns, 
we can extract binary tables with pairs of columns 
$\{ (C_i, C_j) | i, j \in [n], i \neq j \}$,
for a total of $2{{n}\choose{2}}$ such column pairs.
For example, in Figure~\ref{tab:filter}, 
we can conceptually extract all pairs of columns such as
(\texttt{Home Team}, \texttt{Away Team}), 
(\texttt{Home Team}, \texttt{Date}), 
(\texttt{Home Team}, \texttt{Stadium}),
(\texttt{Home Team}, \texttt{Location}), etc.



Because not all these
pairs are meaningful mappings, we filter out candidates 
with a coherence-based filtering 
and a local FD based filtering.

\begin{table}
    \centering
\hspace{-3mm}
\includegraphics[height=0.88in]{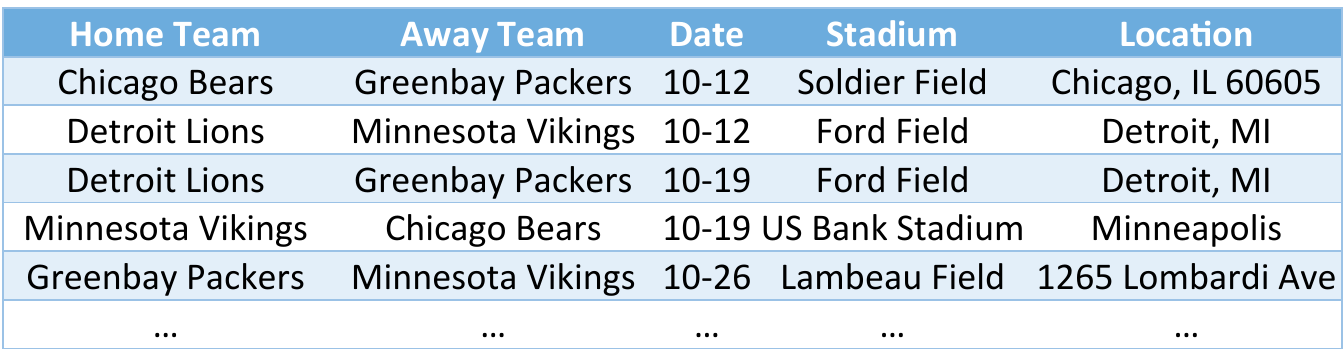}
\vspace{-2.0ex}
\caption{An example input table. Candidate two-column tables can 
be extracted using both PMI and FD filtering.}\label{tab:filter}
\end{table}

\subsection{Column Filtering by PMI}

When given a large table corpus (especially web tables), 
some tables are inevitably of low quality. 
Quality issues can arise because (1) columns may be mis-aligned due
to extraction errors (especially for complicated tables like pivot table and
composite columns); or (2) some table columns just have
incoherent values.

In both of these cases, the resulting table column will appear 
to be ``incoherent'' when looking at all values in this column.
For example, the last column \texttt{Location} in Table~\ref{tab:filter}
have mixed and incoherent values. We would like to exclude such columns
from consideration for mapping synthesis.

Therefore we measure the coherence of a table column 
based on semantic coherence between pairs of values.
We apply a data-driven approach to define coherence 
based on co-occurrence statistics in a
corpus. Let $s(u, v)$ be the coherence between two
values $u$ and $v$. Define
$\mathcal{C}(u) = \{ C | u \in C, C \in T, T \in \mathcal{T} \}$ as the columns
in the table corpus $\mathcal{T}$ containing value $u$, and
define $\mathcal{C}(v)$ similarly. Clearly, if
$\mathcal{C}(u) \cap \mathcal{C}(v)$ is a large set,
it means $u$ and $v$ are co-occurring frequently
(e.g., $u = $ \texttt{USA} and $v = $ \texttt{Canada}). Then
they intuitively are highly related and thus should 
have a high semantic coherence score. 

We use Point-wise Mutual Information (PMI)~\cite{Church:1990} to 
quantify the strength of co-occurrence as a proxy for coherence.
\begin{equation}
\label{eqn:pmi}
\text{PMI}(u, v) = \log \frac{p(u, v)}{p(u)p(v)}
\end{equation} 

Where $p(u)$ and $p(v)$ are the probabilities of seeing 
$u$ and $v$ from a total of $N$ columns in a table corpus
$\mathcal{T}$, defined as 
$p(u)=\frac{|\mathcal{C}(u)|}{N}$, 
$p(v)=\frac{|\mathcal{C}(v)|}{N}$  
and $p(u, v)=\frac{|\mathcal{C}(u) \cap \mathcal{C}(v)|}{N}$. 

\iftoggle{fullversion}
{
\begin{nexp}
Let $u = $\texttt{USA} and $v = $ \texttt{Canada}. 
Suppose $N = 100M$ (there are a total of 100M columns), 
$|\mathcal{C}(u)| = 1000$, $|\mathcal{C}(v)| = 500$, and
$|\mathcal{C}(u) \cap \mathcal{C}(v)| = 300$ 
(individually, the two strings occur 1000 and 500 times
respectively; together they co-occur 300 times). It can be calculated
that PMI$(u, v) = 4.78 > 0$, suggesting that they have high
co-occurrence and strong semantic coherence.
\end{nexp}
}
{
}

We define coherence of two values, denoted by $s(u, v)$, 
as a normalized version of PMI called Normalized PMI (NPMI),
which has a range of $[-1,1]$:
\[ s(u, v) = \text{NPMI}(u, v) = \frac{\text{PMI}(u, v)}{- \log p(u, v)} \]

Using $s(u, v)$, the \textit{coherence score} of a column 
$C= \{ v_1, v_2, ... \}$, denoted as $S(C)$, 
is simply the average of all pair-wise scores.
\begin{equation}
\label{eqn:coherence}
S(C)=\frac{\sum_{v_i, v_j \in C, i<j} s(v_i, v_j)}{{{|C|}\choose{2}}}
\end{equation}

We can then filter out a column $C$ if its coherence $S(C)$ 
is lower than a threshold.


\begin{example}
\label{ex:pmi}
Table~\ref{tab:filter} is an example table with five columns. 
Column coherence computed using NPMI 
in Equation~\eqref{eqn:coherence} would reveal that
the first four columns all have high coherence scores,
because values in these columns co-occur often in the
table corpus.

The last column \texttt{Location}, however, has low
coherence, because values in this column are mixed
and do not co-occur often enough in other columns.
We will remove this column when generating
column pairs.
\end{example}


\subsection{Column-Pair Filtering by FD}

After removing individual columns with low coherence scores, 
we use the resulting table 
$T = \{C_1, C_2, ..., C_n \}$ to generate binary tables with ordered column 
pairs $B(T) = \{ (C_i, C_j) |$ $ i, j \in [n], i \neq j \}$
as candidate tables. 
However, most of these 
two-column tables do not express meaningful mapping relationships,
such as (\texttt{Home Team}, \texttt{Away Team}), 
and (\texttt{Home Team}, \texttt{Date})
in Table~\ref{tab:filter}.

Since our goal is to produce mapping relationships,
we apply local FD checking to prune away column pairs
unlikely to be mappings. As discussed
in Definition~\ref{def:approx} we account for name ambiguity
(like (\texttt{Portland} $\rightarrow$ \texttt{Oregon})
and (\texttt{Portland} $\rightarrow$ \texttt{Maine})) by
allowing approximate FD that holds for 95\% of values.

\begin{example}
Continue with Example~\ref{ex:pmi}, we have pruned
away the last column \texttt{Location} from
Table~\ref{tab:filter} based on coherence scores.
Four columns remain, for a total
of $2{{4}\choose{2}} = 12$ ordered column pairs.
Only $2$ out of the $12$ column pairs satisfy
FD, namely, 
(\texttt{Home Team}, \texttt{Stadium})
and (\texttt{Stadium}, \texttt{Home Team}).
\end{example}



We note that around $78\%$ candidates can be filtered out with these methods.
The procedure used in this 
step can be found in Appendix~\ref{apx:filter}.



\section{Table Synthesis} \label{sec:partition}

Using candidate two-column tables produced from the previous step,
we are now ready to synthesize relationships. Recall that synthesis
provides better coverage for instances (e.g., synonyms) as discussed in the introduction.

\subsection{Compatibility of Candidate Tables} \label{sec:graph}

In order to decide what candidate tables should be stitched together
and what should not, we need to reason about
compatibility between tables.

\begin{figure}[t]
\vspace{-3mm}
\centering
\includegraphics[width=.6\columnwidth]{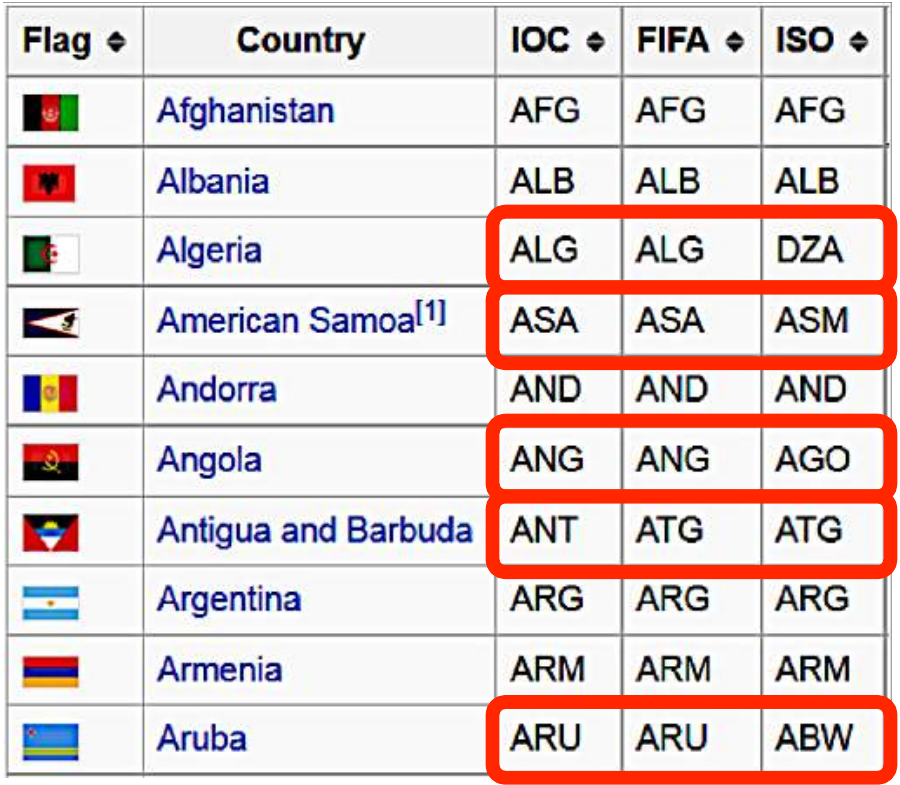}
\caption{Mappings from 
Wikipedia\protect\footnotemark~for country
names and three types of country codes: \texttt{IOC}, 
\texttt{FIFA}, and \texttt{ISO}. The three have
identical codes for many countries, but also different
ones for many others (in red circles).}\label{fig:countrycode}
\end{figure}

\footnotetext{\scriptsize \url{https://en.wikipedia.org/wiki/Comparison_of_IOC,_FIFA,_and_ISO_3166_country_codes}}

\begin{table}
\vspace{-1mm}
    \centering
	\begin{subfigure}[b]{0.13\textwidth}
    		\centering
		\includegraphics[height=0.85in]{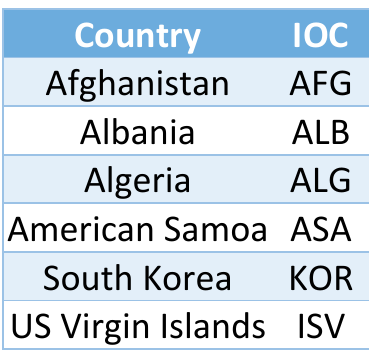}
		\caption{$B_1$: IOC-(1)} \label{tab:ioc1}
	\end{subfigure}
	\begin{subfigure}[b]{0.2\textwidth}
		\centering
		\includegraphics[height=0.85in]{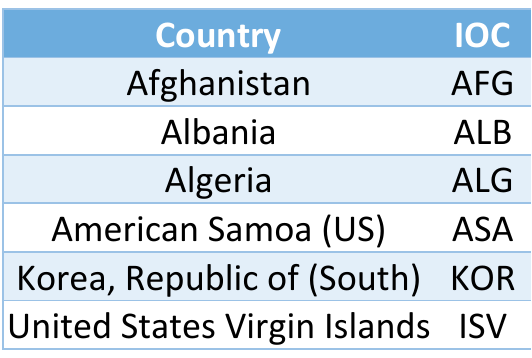}
		\caption{$B_2$: IOC-(2)} \label{tab:ioc2}
	\end{subfigure}
	\begin{subfigure}[b]{0.13\textwidth}
		\centering
		\includegraphics[height=0.85in]{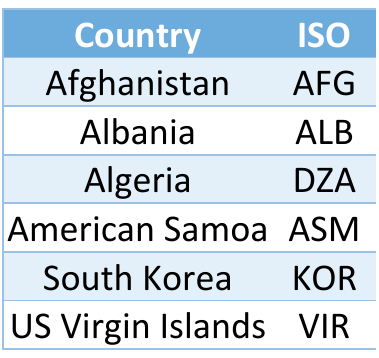}
		\caption{$B_3$: ISO} \label{tab:iso}
	\end{subfigure}
\vspace{-0.2mm}
\caption{Example two-column binary tables for synthesis: 
(a) Countries and IOC codes, 
(b) Countries and IOC codes, where some countries use
alternative synonyms compared to the first table, 
(c) Countries and ISO codes, where
the code for some country can be different from the first two tables.}\label{tab:synthesis}
\end{table}

\paragraph*{Positive Evidence for Compatibility}~\\
Let $B = \{ (l_i, r_i) \}$ and $B' = \{ (l'_i, r'_i) \}$ be
two binary relationships produced by the previous step, 
each with sets of (left, right) value pairs.
If these two relations share many common value
pairs, or $|B \cap B'|$ is large, they
are likely in the same relationship and
compatible for synthesis.

Let $w^+(B, B')$ be the \textit{positive compatibility}
between $B$ and $B'$. We would like to 
use set-based similarity to quantify compatibility
based on the overlap $|B \cap B'|$. 
However, common metrics like Jaccard Similarity, defined as 
$\frac{|B \cap B'|}{|B \cup B'|}$, would not
work because if one small relation is fully contained 
by another ($B \supset B'$, $|B| \gg |B'|$),  the compatibility 
should intuitively be high, but the Jaccard 
Similarity score would actually be low.

Containment metrics would mitigate this issue, 
but Jaccard Containment is asymmetric --
we want it to be symmetric because both the compatibility of $B$, $B'$
and the compatibility of $B'$, $B$ are essentially the same thing 
($w^+(B, B')$ = $w^+(B', B)$). Given these
we use a symmetric variant of Jaccard Containment 
called \textit{Maximum-of-Containment}~\cite{Broder:1997}
for $w^+(B, B')$:
\begin{equation}
\label{eq:positive}
w^+(B, B')
=\max\{\frac{|B \cap B'|}{|B|}, \frac{|B \cap B'|}{|B'|}\}
\end{equation}


\begin{example}
\label{ex:positive}
Table~\ref{tab:synthesis} shows three two-column candidate tables,
$B_1$, $B_2$ and $B_3$, respectively. The first two are for the \texttt{IOC} code, 
while the last is for a different \texttt{ISO} code. All of these three
are valid mappings but are for two different country-code standards, as explained
in Figure~\ref{fig:countrycode}.

Using Equation~\eqref{eq:positive}, we can compute the positive
compatibility between each pair of tables. 
For example, we have $w^+(B_1, B_2) = \max\{\frac{3}{6}, \frac{3}{6}\} = 0.5$,
because $|B_1 \cap B_2| = 3$ (the first three rows), suggesting that
the two tables share a significant fraction of mappings and are likely to
be compatible for synthesis.
\end{example}

\textit{Efficiency.}
Although conceptually compatibility scores can be computed 
for all pairs of candidates, in reality most tables share
no common values, and will have a score of 0.
A practical issue here is that given $N$ total 
candidate tables, we need to perform 
$O(N^2)$ expensive containment computations. 
With millions of tables, this quadratic step
is too expensive even for large Map-Reduce clusters.

In reality we observe that the scores for most pairs of tables are zero since they
share no overlapping values at all. 
For example, Table~\ref{tab:country} is about countries
and Table~\ref{tab:ticker} is about stock tickers. They
have no overlaps in value-pairs, so both 
positive and negative weights are 0. Computing scores for
these non-overlapping sets is clearly wasteful.

To address this problem, we use inverted-index-like regrouping
in a Map-Reduce round to map all tables 
sharing at least some common value-pairs to the same partition,
so that compatibility is computed only for pairs of tables within 
each partition. Specifically,
we evaluate $w^+(B, B')$ only if $B$ and $B'$ share more 
than $\theta_{overlap}$ value pairs (both left and right values), and similarly
we evaluate $w^-(B, B')$ only if $B$ and $B'$ share more 
than $\theta_{overlap}$ left-hand-side values. In practice, the number of 
non-zero weighted edges is much smaller than $N^2$.
This optimization makes it possible to scale the pair-wise computation step 
to hundreds of millions of tables.

\textbf{Approximate String Matching.} In real tables, values from
different tables often have slight variations, 
such as ``Korea, Republic of'' \& 
``Korea Republic'', or ``American Samoa'' \&
``American Samoa (US)''. In practice, there are other
extraneous information in table cells, such as the
footnote mark ``[1]'' in the fourth row
in Figure~\ref{fig:countrycode}.
These artificially reduce positive compatibility and 
in some cases increase
negative compatibility between tables, which is undesirable.

To account for such minor syntactic variations,
we use approximate string matching between cell values.
Specifically, we measure the Edit 
Distance, denoted as $d_{ed}(v_1, v_2)$,
between a pair of values $v_1$ and $v_2$. 
We treat $v_1$ and $v_2$ as a match 
if $d_{ed}(v_1, v_2)$ is smaller than a
threshold $\theta_{ed}$. Here we use a fractional
threshold defined as 
$\theta_{ed} = \min\{\lfloor|v_1| \cdot f_{ed}\rfloor, \lfloor|v_2| \cdot f_{ed}\rfloor\}$, 
which is dynamically determined based on the length
of string $|v_1|$, $|v_2|$, and a fixed fractional
value $f_{ed}$ (e.g., 0.2). 
We choose to use a fractional distance
instead of an absolute distance, because the desired edit distance
should change based on the length of values. 
For example, for short values such as
``USA'' or ``RSA'' (for South Africa), any 
absolute distance threshold $\geq 1$
would incorrectly match the two. Fractional
threshold on the other hand would require 
an \textit{exact} match for short strings like these.
We further restrict the threshold to be
within some fixed threshold $k_{ed}= 10$ to safeguard false positives.
Combining, we use
$\theta_{ed}(v_1, v_2) = \min\{\lfloor|v_1| \cdot f_{ed}\rfloor, \lfloor|v_2| \cdot f_{ed}\rfloor, k_{ed} \}$.

\begin{example}
\label{ex:approx}
We continue with Example~\ref{ex:positive} in
Table~\ref{tab:synthesis}. When using
approximate matching for positive compatibility,
$w^+(B_1, B_2)$ will now be updated to 
$\max\{\frac{4}{6}, \frac{4}{6}\} = 0.67$.
This is because in addition to the first three matching rows 
between $B_1$ and $B_2$, now
the fourth row ``American Samoa'' and
``American Samoa (US)'' will also be considered 
as a match, as the Edit Distance
between the two values is 2 (ignoring
punctuations), which is no greater than
$\theta_{ed} = \min\{\lfloor13 \cdot 0.2\rfloor, \lfloor15 \cdot 0.2\rfloor, 10 \} = 2$.
\end{example}

\textit{Efficiency}. There are hundreds of millions of table pairs 
for which we need to compute compatibility. Let
$m$ and $n$ be the numbers of values in a pair of tables.
For each pair we need to make $O(nm)$
approximate string comparisons, each of which
is in turn $O(|v_1||v_2|)$ when using conventional dynamic programming
on the full matrix.
This is too expensive even for production Map-Reduce clusters.

Our observation is that the required edit distance threshold $\theta_{ed}$ is
small in most cases. So using ideas similar to the 
Ukkonen's algorithm~\cite{Ukkonen:1985}, we only compute DP on 
the narrow band in the diagonal direction of the matrix, which
makes it $O(\theta_{ed} \cdot \min\{|v_1|, |v_2|\})$. 
Since $\theta_{ed}$ is small it makes this step feasible.
Pseudo-code of this step can be found
in Appendix~\ref{apx:approx-string}.

\textbf{Synonyms.} In some cases, synonyms of entity names
may be available, e.g., using existing synonym feeds 
such as~\cite{synonyms}. If we know, for instance,
``US Virgin Islands'' and ``United States Virgin Islands''
are synonyms from external sources, we can 
boost positive compatibility between 
$B_1$ and $B_2$ in Table~\ref{tab:synthesis} accordingly.
We omit discussions on possible lookup-based
matching in the interest of space.

\paragraph*{Negative Evidence for Incompatibility} ~\\
Positive evidence alone is often not sufficient to fully capture 
compatibility between tables, as tables of different
relationships may sometimes have substantial overlap.
For example, it can be computed that the positive
compatibility between $B_1$ in Table~\ref{tab:ioc1}
and $B_3$ Table~\ref{tab:iso} is
$\max\{\frac{3}{6}, \frac{3}{6}\} = 0.5$ 
(the first, second and fifth rows match).
Given the high score, the two
will likely merge incorrectly (note that one 
is for \texttt{IOC} code while the other is for \texttt{ISO}).
This issue exists in general when one of the
columns is short and ambiguous (e.g. codes), or when one
of the tables has mixed values from different mappings
(e.g., both city to state and city to country).

We observe that in these cases the two tables actually also contain
\textit{conflicting} value pairs, such as the third and fourth row 
in the example above 
where the two tables have the same left-hand-side value, but
different right-hand-side values.
This violates the definition of mapping relationship, and
is a clear indication that the two tables are not 
compatible, despite their positive scores.

We thus introduce a negative \emph{incompatibility} between 
tables. Given two tables $B$ and $B'$, define their \textit{conflict set}
as $F(B, B') = \{ l | (l, r) \in B, (l, r') \in B', r \neq r' \}$, or
the set of values that share the same left-hand-side but 
not the right-hand-side. For example, between $B_1$ in Table~\ref{tab:ioc1}
and $B_3$ Table~\ref{tab:iso}, (Algeria, ALG) and (Algeria, DZA) is a 
conflict. 

To model the (symmetric) incompatibility between two tables $B$ and $B'$,
we define a negative incompatibility score $w^-(B, B')$ similar to
positive compatibility in Equation~\eqref{eq:positive}:
\begin{equation}
\label{eq:negative}
w^-(B, B')= - \max\{\frac{|\textit{F}(B, B')|}{|B|}, \frac{|\textit{F}(B, B')|}{|B'|}\}
\end{equation}

\begin{example}
\label{ex:negative}
We continue with Example~\ref{ex:approx} in
Table~\ref{tab:synthesis}. 
As discussed earlier, the positive
compatibility between $B_1$ in Table~\ref{tab:ioc1}
and $B_3$ in Table~\ref{tab:iso} is
$\max\{\frac{3}{6}, \frac{3}{6}\} = 0.5$,
which is substantial and will lead to incorrect merges
between two different relationships (\texttt{IOC} and \texttt{ISO}).

Using negative incompatibility, we can compute
$w^-(B_1, B_3)$ as
$-\max\{\frac{3}{6}, \frac{3}{6}\} =- 0.5$,
since the third, forth and sixth rows conflict
between the two tables, and both tables have 6 rows.
This suggests that $B_1$ and $B_3$ have substantial
conflicts, indicating that a merge will be inappropriate.

In comparison, for $B_1$ in Table~\ref{tab:ioc1}
and $B_2$ in Table~\ref{tab:ioc2}, which talk
about the same relationship of \texttt{IOC},
their conflict set is empty and $w^-(B_1, B_2) = 0$,
indicating that we do not have negative evidence
to suggest that they are incompatible.
\end{example}


\subsection{Problem Formulation for Synthesis}
We use a graph $G = ( \mathcal{B}, E )$ 
to model candidate tables and their 
relationships, where $\mathcal{B}$ is the union of all binary tables produced in 
the preprocessing step in Section~\ref{sec:preprocessing}. 
In $G$ each vertex represents a
table $B \in \mathcal{B}$. Furthermore,
for each pairs of
vertices $B, B' \in \mathcal{B}$,
we use compatibility scores $w^+(B, B')$
and incompatibility scores $w^-(B, B')$
as the positive and negative edge weights of the graph.

\begin{figure}
\centering
\includegraphics[width=1.\columnwidth]{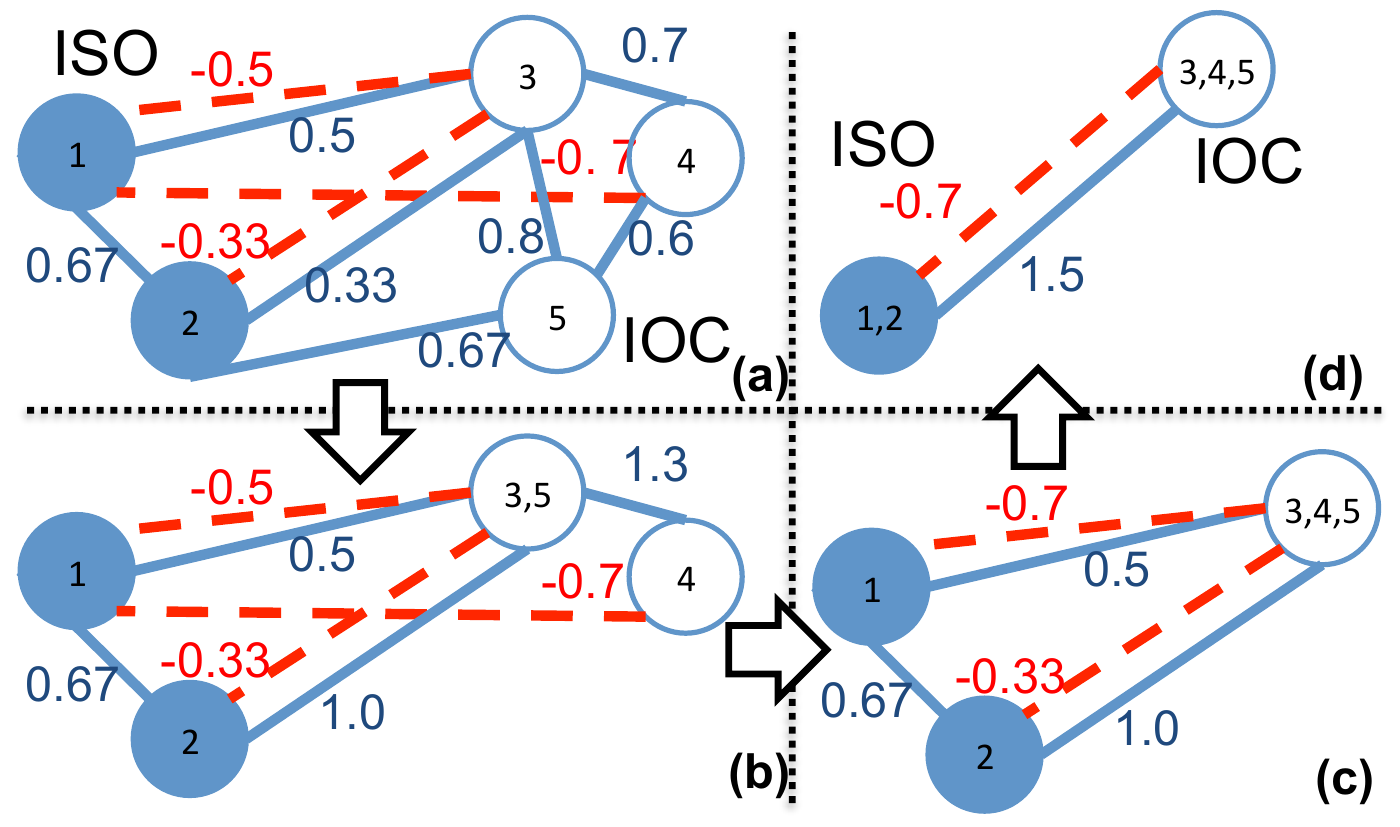}
\caption{Graph representation of candidate tables. 
Solid vertices on the left represent tables for ISO codes;
hollow vertices on the right represent tables for IOC codes. 
Furthermore, solid edges indicate positive compatibility, 
while dashed edges indicate negative incompatibility.
Edges with weight of 0 are omitted on the graph.}\label{fig:partitioning}
\end{figure}

\begin{example}
\label{ex:graph}
Given the tables $B_1$, $B_2$ and $B_3$ in
Table~\ref{tab:synthesis}, we can represent
them and their compatibility relationships
as a graph as in Figure~\ref{fig:partitioning}(a).

As discussed in Example~\ref{ex:approx}, the positive
compatibility between $w^+(B_1, B_2) = 0.67$, which
is shown as solid edge with positive weight in this graph.
Similarly we have negative edge weights like 
$w^-(B_1, B_3) = -0.5$ as discussed
in Example~\ref{ex:negative}. This graph
omits edges with a weight of 0, such as 
$w^-(B_1, B_2)$.
\end{example}

Since we need to synthesize compatible
tables into larger mapping relationships, in the context of graph $G$ 
we need to group compatible vertices/tables 
together. This naturally corresponds to a partitioning 
$\PartitionSet=\{P_1, P_2, ...\}$ of 
$\mathcal{B}$, where each $P_i \subseteq \mathcal{B}$ represents
a subset of tables that can be synthesized into
one relationship. Since different partitions 
correspond to distinct relationships,
the partitioning should be disjoint ($P_i \cap P_j = \emptyset,  i \neq j$),
and they should collectively cover $\mathcal{B}$, or
$\bigcup_{P \in \PartitionSet}{P} = \mathcal{B}$. 

Intuitively, there are many ways to partition
$\mathcal{B}$ disjointly, but we
want to find a good partitioning that has
the following desirable properties: (1) compatible
tables are grouped together as much as 
possible to improve coverage of individual mapping relationships;
and (2) incompatible tables should
not be placed in the same partition.

We translate these intuitive requirements into
an optimization problem. First, we want
each partition $P$ to have as many
compatible tables as possible. Let $w^+(P)$ be
the sum of positive compatibility in a partition $P$:
$$w^+(P) = \sum_{B_i, B_j \in P, i<j} w^+(B_i, B_j)$$
We want to maximize the sum of
this score across all partitions, or $\sum_{P \in \PartitionSet}{w^+(P)}$.
This is our optimization objective.

On the other hand, we do not want to put incompatible
tables with non-trivial $w^-$ scores, 
such as $B_1$ and $B_3$ in Example~\ref{ex:graph},
in the same partition. Since we disallow this to happen,
we treat edges with negative scores $w^-$ below a
threshold $\tau$ as \textit{hard-constraints}. Note that a negative
threshold $\tau$ (e.g., $-0.2$) is used in place of 0 
because we do not over-penalize tables with
slight inconsistency due to minor
quality and extraction issues. We ignore
the rest with insignificant negative scores by essentially
forcing them to 0.
Let $w^-(P)$ be the sum of substantial negative weights in $P$ defined below.
$$w^-(P) = \sum_{B_i, B_j \in P, w^-(B_i, B_j) < \tau} w^-(B_i, B_j)$$
We use this as a constraint of our formulation -- we want no edges in the
same partition to have substantial conflicts, or,
$w^-(P) = 0, \forall P \in \mathcal{P}$.

Putting these together, we formulate table synthesis 
as follows.

\begin{problem}[Table Synthesis] \label{problem:partitioning}
\begin{align}
\mbox{max} \qquad & \sum_{P \in \PartitionSet}{w^+(P)} \label{eq:obj} \\
 \mbox{s.t.} \qquad 
 & \sum_{P \in \PartitionSet}{w^-(P)} = 0 \label{eq:conflict} \\
& P_i \bigcap P_j = \emptyset, \quad~\quad \forall P_i \neq P_j  \label{eq:partition1} \\
 & \bigcup_{P \in \PartitionSet} P = \mathcal{B} \label{eq:partition2}
\end{align} 
\end{problem}
By placing compatible tables in the same partition, we 
score more in the objective function in Equation~\eqref{eq:obj},
but at the same time
Equation~\eqref{eq:conflict} guarantees that no conflicting
negative edge can be in the same partition.
Equation~\eqref{eq:partition1} and~\eqref{eq:partition2} are
used to ensure that
$\PartitionSet$ is a proper disjoint partitioning.

\begin{example}
We revisit the example in Figure~\ref{fig:partitioning}(a).
Using the formulation above, it can be verified that 
the best partitioning is 
$\{\{B_1, B_2\},$ $\{B_3, B_4, B_5\}\}$, which
groups two ISO tables and three IOC tables into separate partitions. 
This partitioning has a total score of 
$2.77$  based on
Equation~\eqref{eq:obj}, without violating constraints
in Equation~\eqref{eq:conflict} by not placing negative
edges in the same partition.

It is worth noting that existing techniques like
schema matching~\cite{Rahm01} only consider
positive similarity (because FD do not generally hold
in tables), and as a result merge
all 5 tables in this example, producing results of low quality.
\end{example}

\begin{thm} \label{1}
The problem~\textit{Table-Synthesis} is NP-hard.
\end{thm}

We prove this using a reduction from graph multi-cut. 
There also exists a trichotomy of
complexity depending on the number of negative edges in the graph.
A proof of the hardness can be found in Appendix~\ref{apx:hardness}.

\iftoggle{fullversion}{
What is interesting is that there exists a trichotomy result
in terms of complexity~\cite{Demaine:2006}. Specifically,
if the graph has exactly 1 negative edge, the problem 
is equivalent to min-cut, max-flow because we can make the pair
of vertices incident to the negative edge as source and sink,
respectively. When there exist 2 negative 
edges, the problem can be solved in polynomial 
time using results from~\cite{Yannakakis:1983}. In the 
more general case when there are 
no fewer than $3$ negative edges, the
problem becomes NP-hard.

Despite the hardness, there is a 
$O(\log N)$-approximation algorithm
for the loss-minimization version of 
Problem~\ref{problem:partitioning}.
Specifically, the loss-minimization version 
of the problem can be written as follows,
which minimizes the positive edge weights
that are lost as a result of the partitioning that
disconnects all the negative edges.

\begin{problem}[Loss Minimization] \label{problem:loss}
\begin{align}
\mbox{min} \qquad & \sum_{B \in P_i, B' \in P_j, i \neq j}{w^+(B, B')} \label{eq:loss-obj} \\
 \mbox{s.t.} \qquad 
 & \sum_{P \in \PartitionSet}{w^-(P)} = 0 \label{eq:loss-conflict} \\
& P_i \bigcap P_j = \emptyset, \quad~\quad \forall P_i \neq P_j  \label{eq:loss-partition1} \\
 & \bigcup_{P \in \PartitionSet} P = \mathcal{B} \label{eq:loss-partition2}
\end{align} 
\end{problem}

Using standard embedding techniques, we can encode
partition decisions using distance variables $d_{ij}$.
$d_{ij}=0$ if vertices $B_i$ and $B_j$ are in the same partition,
and $d_{ij}=1$ if they are in different partitions. This produces
the following formulation.

\begin{problem}[Embedding] \label{problem:ILP}
\begin{align}
\mbox{min} & \quad \sum{ w^+(B_i,B_j) \cdot d_{ij} } \notag \\
\mbox{s.t.}  & \quad{} d_{ij} + d_{jk} \ge d_{ik}, \quad{} \forall i, j, k \label{eq:ILPpartition1} \\
 & \quad{} d_{ij} = d_{ji} \label{eq:ILPpartition2} \\
 & \quad{} d_{ij} \in \{0, 1\} \\
 & \quad{} d_{ij} = 1, \qquad{} \quad~\quad \forall w^-(B_i, B_j) < \tau \label{eq:ILPconflict}
\end{align}
\end{problem}

This problem is known to be APX-hard~\cite{Dahlhaus:1994}.
Using classical techniques including LP-relaxation and 
region-growing for randomized 
rounding~\cite{Bejerano06, garg1994multiway, varizani}, it is possible to produce 
$O(\log N)$ approximation. More discussions of this can be found 
in Appendix~\ref{apx:lp} in the interest of space.

Such an approximation scheme requires to model each pair
of vertices as a decision variable $d_{ij}$, and then solve the associated
LP before applying randomized rounding. While it may be practical
for problems of moderate sizes, we are dealing with graphs with
millions of vertices, where solving an LP with 
a quadratic number of variables is clearly infeasible.
} 
{
We can show that the loss-minimization version of this problem can
be solved using LP-relaxation and randomized 
rounding~\cite{Bejerano06}, to produce $O(log N)$ approximation.
Details of this LP-based solution can be found in a full version of this paper.
While this LP based solution is practical for problems of moderate sizes, 
we are dealing with graphs with
millions of vertices and a quadratic number of variables, for which
existing LP-solvers cannot currently handle.
} 

As a result, we use an efficient heuristic to 
perform greedy synthesis. 
Specifically, 
we initially treat each vertex as 
a partition.
We then iteratively merge a pair of partitions $(P_1, P_2)$ that are 
the most compatible
to get a new partition $P'$, and update the remaining
positive/negative edges.
The algorithm terminates when no partitions can be merged. 
Pseudo-code of this procedure can be found in Appendix~\ref{apx:table-partition-algo}.

\textit{Efficiency.}
While the procedure above appears
straightforward for graphs that fit in a single machine, scaling to large graphs on 
Map-Reduce is not straightforward.
We use a divide-and-conquer approach to first produce
components that are connected non-trivially by positive edges
on the full graph, and then look at each subgraph individually.
More discussions on this step can be found in
Appendix~\ref{apx:connected}.

\begin{example}
Figure~\ref{fig:partitioning} shows how 
Algorithm~\ref{alg:greedy} works on a small graph. 
The algorithm first merges $\{B_3\}$ and $\{B_5\}$ to 
get Figure~\ref{fig:partitioning}b because Edge 
$(\{B_3\}, \{B_5\})$ has the greatest weight. The weight 
of Edge $(\{B_2\}, \allowbreak \{B_3, B_5\})$ changes as 
$w^+(\{B_2\},\allowbreak \{B_3, B_5\}) \Leftarrow w^+(\{B_2\}, \{B_3\}) + w^+(\{B_2\}, \{B_5\})$. The weight of Edge $(\{B_4\}, \allowbreak \{B_3, B_5\})$ 
also changes similarly.

The algorithm then merges $\{B_3, B_5\}$ and $\{B_4\}$ to get Figure~\ref{fig:partitioning}c and finally combines $\{B_1\}$ and $\{B_2\}$ to get Figure~\ref{fig:partitioning}d. The algorithm stops because of the negative weight between $\{B_1, \allowbreak B_2\}$ and $\{B_3, B_4, B_5\}$.
\end{example}


\paragraph*{Conflict Resolution}
\vspace{-3mm}
\begin{figure}[t]
\centering
\includegraphics[width=1\columnwidth]{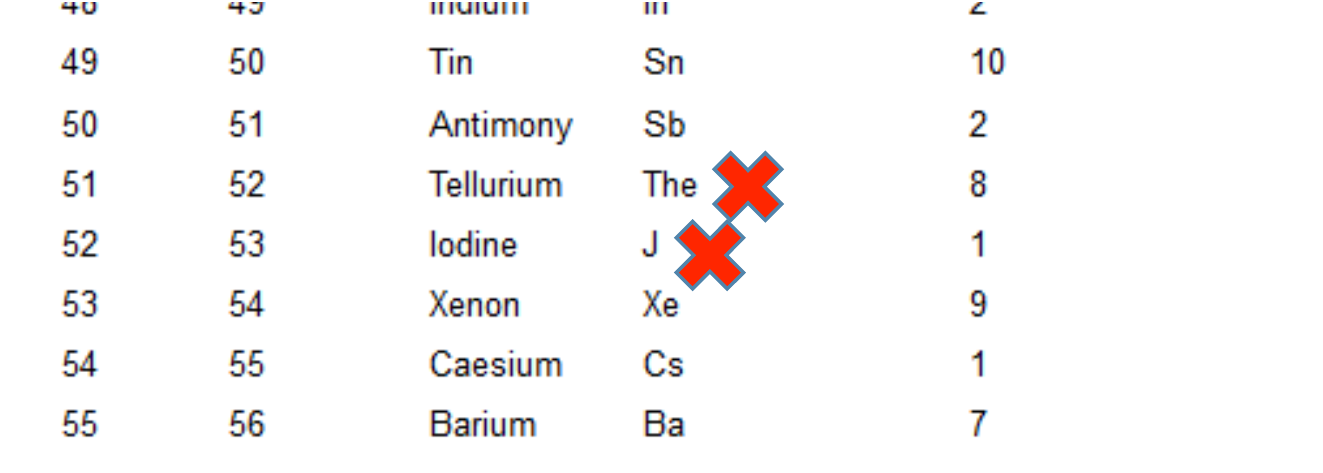}
\caption{A real table with errors that can cause conflicts.}\label{fig:truth}
\end{figure}

We observe that synthesized relations often have
conflicts that require post-processing. 
Specifically, when we union all tables in the same partition together,
there will be a small fraction of rows that share the same left-hand-side
value, but have different right-hand-side values. 
This could be due to quality issues in the original input tables,
such as the example in Figure~\ref{fig:truth} that has 
incorrect chemical symbols for two of the rows 
(the symbol of \texttt{Tellurium} should be \texttt{Te} and
\texttt{Tellurium} should be \texttt{I}). 
Quality issues like this are
actually common in large corpus,
and manifest themselves as inconsistent mappings in synthesized results.
Since the majority of tables in the partition 
should agree with the ground-truth mapping,
we resolve conflicts by removing the least
number of low-quality tables, such that the resulting partition has
no conflicts.

Let $P$ be a partition with 
candidate tables $\{B_1, B_2, ...\}$, each of which is 
a set of value pairs $B_i = \{(l, r)\}$. 
Recall that in Section~\ref{sec:graph} we define 
a \textit{conflict set} $F(B, B')$ 
to be $\{ l | (l, r) \in B, (l, r') \in B', r \neq r' \}$.
We can again leverage synonyms and do not
treat $(l, r), (l, r')$ as conflicts if $(r, r')$ are known
to be synonyms. 

Now we want to find out the largest subset 
$P_T \subseteq \mathcal{P}$ such that no two tables in $P_T$ 
conflict with each other, which can be formulated as follows.

\begin{problem}[Conflict Resolution] \label{problem:truth}
\begin{align}
\mbox{max} \qquad & \left\vert \bigcup_{B_i \in P_T} B_i \right\vert \notag \\
 \mbox{s.t.} \qquad & F(B_i, B_j) = \emptyset, \quad \forall B_i, B_j \in P_T \label{eq:truthConflict}
\end{align} 
\end{problem}
The objective is to include as many value pairs as possible, under
the constraint that no pairs of tables in the selected subset $P_T$ can have conflict.


This problem is NP-hard (reduction from Independent Set).
So we iteratively find and remove a value pair that conflicts with 
the most other value pairs. 
Pseudo-code of this procedure can be found
in Appendix~\ref{apx:postprocessing}.

\iftoggle{fullversion}
{
\paragraph*{Table Expansion}
Another potential issue is that for large mapping
relationships such as (\texttt{airport-name}, \texttt{airport-code})
that has more than 10K instances, synthesized tables 
may still miss values that are unpopular 
with little or no presence in web tables. 
We expand synthesized results that are robust ``cores'',
using trustworthy sources such as data.gov or spreadsheet 
files (\emph{.xlsx}).
Appendix~\ref{apx:table-expansion} gives more details for this
optional step.
}
{
}

We note that there are many existing methods for conflict 
resolution~\cite{Li16} that can conceptually be applied to 
the post-processing step, and it is interesting to explore 
their applicability. Because we do not consider this post-processing 
step to be our key contribution, and we include this step here 
for completeness, we do not perform an exhaustive comparison.

\subsection{Synthesized Mappings for Curation}
\label{sec:curation}
While the synthesized mappings produced by our algorithm
are generally of high quality, for many applications 
a very high precision is required. For example, for commercial
spreadsheet software like Excel, any error 
introduced by black-box algorithms can be hard to detect by
users, but has
damaging consequences and thus unacceptable. 
In such settings, our approach of pre-computing all
candidate mappings from table corpora allows humans to
inspect and curate these mappings to ensure
very high accuracy. High-quality mappings produced by
automatic algorithms can greatly reduce the effort
required by human curators.

It is interesting to note that synthesized results we
produce have a natural notion of importance/popularity.
Specifically, for each synthesized mapping, we 
have statistics such as the number of web domains
whose tables contributed to this mapping, and how
many raw tables are synthesized in the same cluster, etc.
Such statistics are very well correlated to the importance
of the mapping, because the more it occurs in the table
corpus, the more likely it is frequently used and important.
This property makes results produced by our approach
amenable to human curation -- instead of looking at a
full corpus with millions of tables, one just needs to look
at synthesized results popular enough.

\iftoggle{fullversion}{
In our experiments using a web corpus, 
we only use about 60K synthesized 
mappings from at least 8 
independent web domains, which is 
orders of magnitude less than
the the number of input tables. Additional 
filtering can be performed to further
prune out numeric and temporal relationships.

While most mappings are static that rarely change, some are 
temporal in nature and may be changing over time. But like 
knowledge-bases used by search engines that also face the 
same ``data freshness'' problem (e.g., when a famous actor gets 
newly married, the knowledge-card used by search engines should 
reflect that new fact within a short period of time), algorithms and 
human curation can for the most part mitigate this problem (e.g., 
regularly refreshing the data by rerunning the pipeline and alert human 
curator for changes). Additional mechanisms include crowd-sourcing that 
allows users to report/flag stale values for them to be corrected. We would 
like to note that because a large fraction of the mappings  harvested are static
 in nature, a one-shot curation of just these mappings can already produce 
 significant values to a variety of applications.

} 
{
} 


\section{Experiments} \label{sec:evaluation}

\subsection{Experimental Setup}

\paragraph*{Table Corpus}
We use two table corpora for our evaluation.

The first table corpus, henceforth denoted as \web,
has over 100 million tables crawled and extracted
from the public web.  
These tables cover diverse domains of interests.

The second table corpus, denoted as \ent, 
has about 500K tables extracted from spreadsheets files
crawled from the intranet of a large IT company.

\paragraph*{Computing Environment}
We implemented algorithms described
in this paper as Map-Reduce programs.
We ran our jobs in a large Map-Reduce 
cluster, alongside with
other production jobs. 
Our input  for \web~has about 223M two-column tables with a size of over 200GB.


\paragraph*{Benchmarks}
We have built a benchmark dataset to evaluate our framework on \web~table corpus~\footnote{\small Mappings in the web benchmark is available at \url{https://www.microsoft.com/en-us/research/publication/synthesizing-mapping-relationships-using-table-corpus/}}. This benchmark dataset contains 80 desirable mapping relationships that we manually curated. These relationships are collected from two sources.

\begin{figure}[t]
\begin{center}
\begin{tabular}{|l|l|}
\hline
\small list of countries and capitals & \small list of pokemons and categories \\ \hline
\small list of car models and makes & \small list of amino acids and symbols \\ \hline
\end{tabular}
\end{center}
\caption{Example queries with ``list of A and B''} \label{tab:queryExample}
\end{figure}%

\begin{itemize}[noitemsep, nolistsep, leftmargin=*, wide=0\parindent]
\item{Geocoding:} 
We observe that geography is a common domain with rich mapping
relationships that are often used in auto-join and auto-correction 
scenarios. Examples here include geographical and administrative coding such
as country code, state code, etc. So we take 14 cases from a Wikipedia list 
of geocoding systems\footnote{\small \url{https://en.wikipedia.org/wiki/Geocoding}}. We omit codes that are impossible to enumerate such as military grid reference system, 
and ones not completely listed on Wikipedia such as HASC code. Figure~\ref{tab:geocoding} lists all cases we take.
\item{Query Log:}
We sample queries of the pattern ``list of A and B''
in Bing query logs that search for mapping relationships.
Figure~\ref{tab:queryExample} shows a few examples with true mappings.
\end{itemize}

\begin{figure}[t]
\begin{center}
\begin{small}
\begin{tabular}{|l|l|}
\hline
FIPS 5-2 & ISO 3166-1 Alpha-3 \\ \hline
FIPS 10-4 & ISO 3166-1 Numeric \\ \hline
IANA Country Code & ITU-R Country Code \\ \hline
IATA Airport Code & ITU-T Country Calling Code \\ \hline
ICAO Airport Code & MARC Country Code \\ \hline
IOC Country Code & NUTS (EU) \\ \hline
ISO 3166-1 Alpha-2 & SGC Codes (Canada) \\ \hline
\end{tabular}
\end{small}
\end{center}
\caption{Geocoding Systems} \label{tab:geocoding}
\end{figure}%

For \web, after selecting mapping relationships,
we curate instances for each relationship, by combining
data collected from web tables as well as knowledge bases.
Specifically, we find a group of tables for each relationship, and then
manually select high-quality ones to merge into the ground truth. 
Finally we combine these high-quality web tables with instances in 
Freebase and YAGO if they have coverage. Note that the resulting
mapping relationships have rich synonyms for the same
entity (e.g., as shown in Table~\ref{tab:synonym}), as well as more
comprehensive coverage for instances.
Constructing such a benchmark set, and ensuring its
correctness/completeness is a 
time-consuming process. 
We intend to publish this benchmark set online
to facilitate future research in this area.

\ent~is more difficult to benchmark because of the difficulty in
ensuring completeness of instances in certain mappings
-- the ground truth may be in master
databases for which we have no access 
(e.g., \texttt{employee} and \texttt{login-alias}). 
Nevertheless, we built 30 best effort benchmark cases.
Recall results on these tests should be interpreted
as relative-recall given the difficulty to ensure completeness. 

\paragraph*{Metrics} We use the standard precision, recall and f-score to measure the performance. Let $B^* = \{(l^*, r^*)\}$ be a ground truth mapping,
and $B =\{(l, r)\}$ be a synthesized relationship for which we want to
evaluate its quality. The precision of $B$ is defined as $\frac{|B \cap B^*|}{|B|}$, the 
recall is $\frac{|B \cap B^*|}{|B^*|}$, and the 
f-score is $\frac{2\text{precision} \cdot \text{recall}}{\text{precision} + \text{recall}}$.

\paragraph*{Methods compared}
We compare the following methods.
\begin{itemize}[noitemsep, nolistsep, leftmargin=*, wide=0\parindent]
\item{\texttt{UnionDomain}.} Ling and Halevy et al.~\cite{Ling:2013} propose to union 
together tables within the same website domain, if their column 
names are identical but row values are disjoint. 
We apply this technique by essentially grouping tables based on
column names and domain names. We evaluate the resulting
union tables against each benchmark case by picking the union table
with the highest F-score.

\item{\texttt{UnionWeb}.} Noticing that only union-ing tables in the
same domain may be restrictive and missing instances for large
relationships, we extend the previous approach to also merge tables
with the same column names across the web, and evaluate all benchmark cases
like above. This is a variant of \texttt{UnionDomain}.

\item{\texttt{Synthesis}.} 
This is our approach 
that synthesizes mapping relationships
as described in Section~\ref{sec:partition}. 

\item{\texttt{SynthesisPos}.} This is the same as \texttt{Synthesis}
except that it does not use the negative signals induced by FDs. 
This helps us to understand the usefulness
of negative signals.

\item{\texttt{WiseIntegrator~\cite{He04, He03}}.} This is a 
representative method in a notable branch in schema matching that collectively
matches schemas extracted from Web forms. It measures the similarity between candidates 
using linguistic analysis of attribute names and value types, etc., and performs 
a greedy clustering to group similar attributes.

\item{\texttt{SchemaCC}.} 
In this method, we mimic pair-wise schema matchers that use the same positive/negative similarity as our approach. Because match decisions are pair-wise, we aggregate these to a
group-level based on transitivity (e.g., if table A matches B and B matches C, then A also matches C). This is implemented as connected components on very large graphs, where
edges are threshold based on a weighted combination of positive/negative scores. We tested different thresholds in the range of $[0, 1]$ and report the best result.

\item{\texttt{SchemaPosCC}.} This is the same as \texttt{SchemaCC} but without
negative signals induced by FDs, since they are not explored in the schema matching literature. We again test thresholds in $[0, 1]$ and report the best number.

\item{\texttt{Correlation~\cite{Chierichetti14}}.} 
In this method, we again mimic pair-wise schema matchers with the same positive/negative 
scores as \texttt{Synthesis}. Instead of using connected components for aggregation as in \texttt{SchemaCC} above, here we instead use the correlation clustering that handles graphs with both positive or negative weights. We implement the state-of-the-art correlation clusterin on map-reduce~\cite{Chierichetti14}, which requires $O(\log|V| \cdot \Delta^+)$ iterations and takes a long time to converge ($|V|$ is the number of vertices of the graph and $\Delta^+$ is the maximum degree of all vertices). We timeout after $20$ hours and evaluate the results at that point.


\item{\texttt{WikiTable}.} Wikipedia has many high-quality tables covering various domains, many of which have mapping relationships. 
To understand the quality of using raw tables 
instead of performing synthesis, we also evaluate each benchmark 
case by finding best pair of columns in a Wikipedia table
that has the highest in F-score.

\item{\texttt{WebTable}.} This method is very similar to
the previous \texttt{WikiTable}, but use all tables 
in the \web~corpus instead of just Wikipedia ones. 

\item{\texttt{Freebase}.} Freebase~\cite{Bollacker:2008} is a well-known 
knowledge base that has been widely used. We obtained its RDF 
dump\footnote{\small \url{https://developers.google.com/freebase/}} and extract 
relationships by grouping RDF triples by their predicates. 
We treat the subject $\rightarrow$ object as one candidate relationship, 
and the object $\rightarrow$ subject as another candidate.

\item{\texttt{YAGO}.}  YAGO~\cite{Suchanek:2007} is another 
public knowledge base that is extensively used. We process a 
YAGO data dump similar to Freebase, by grouping
YAGO RDF triples using their predicates to form subject-object 
and object-subject relationships.

\end{itemize}

Note that in all these cases, we score each benchmark case by
picking the relationship in each data set that has the best f-score.
This is favorable to all the methods -- a human who 
wishes to pick the best relationship to be 
used as mappings, and who could afford to inspect all these tables, would
effectively pick the same tables.

\begin{figure}[]
\centering
\includegraphics[width=1.\columnwidth]{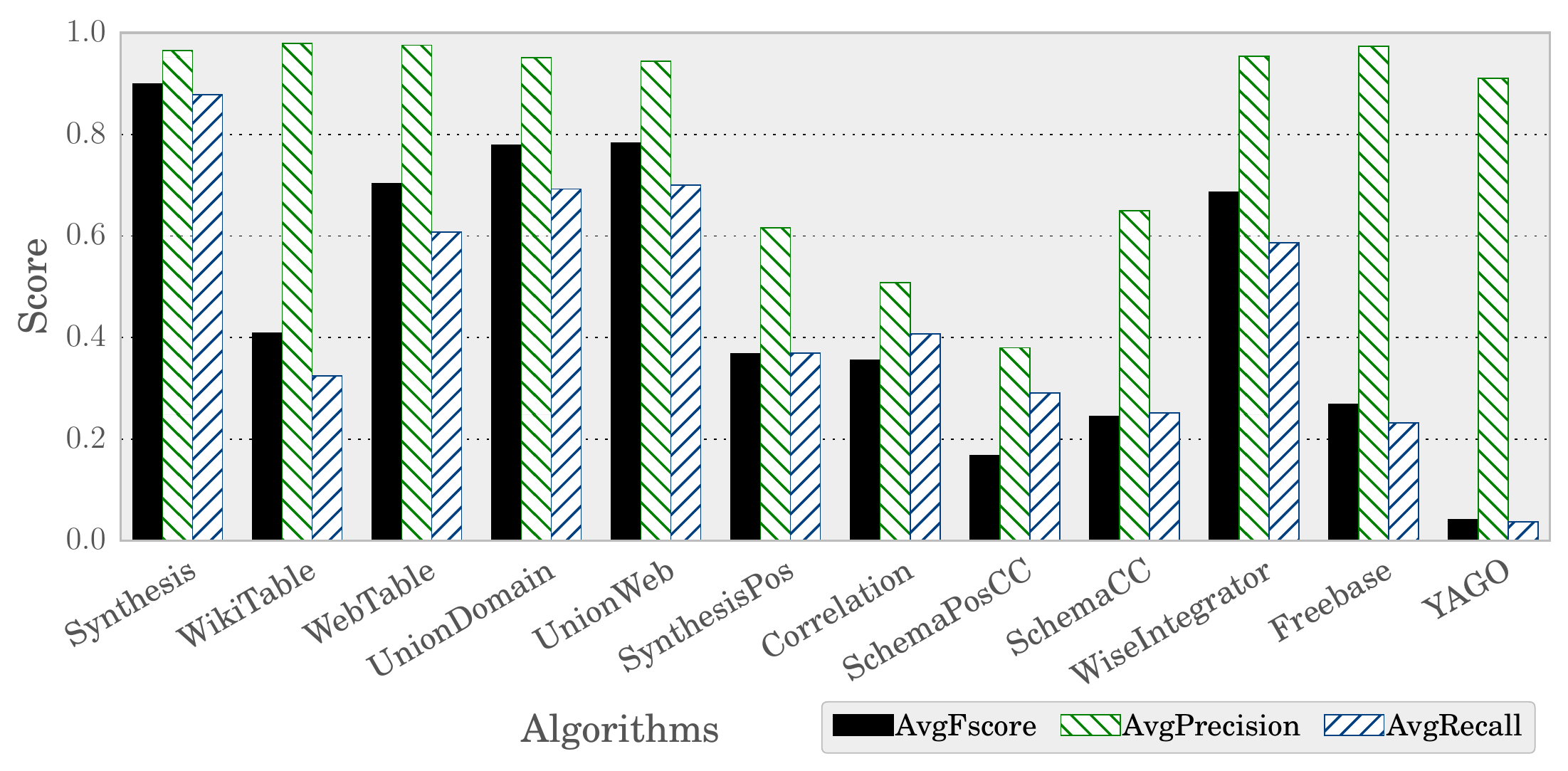}
\caption{Average f-score, precision and recall comparison.} \label{fig:baseline}
\end{figure}

\subsection{Quality Comparison}

Figure~\ref{fig:baseline} shows the average f-score, precision and recall
across all 80 benchmark cases in the \web{} benchmark for all methods compared. 
\texttt{Synthesis} scores the best in average recall (0.88) and f-score (0.90), while \texttt{WikiTable} has the best 
average precision (0.98)~\footnote{\small Since \texttt{WikiTable}
methods miss many relationships, we exclude cases whose precision
is close to 0 from the average-precision computation. 
This makes the average precision favorable to \texttt{WikiTable}.
The same is applied to other table and knowledge based methods.}. 
%

In comparison, using only raw tables from \texttt{WikiTable} with no synthesis
has high precision
but low recall, because not only are certain
instances and synonyms missing (these tables tend to be short
for human consumption), many relationships are also missing
altogether from \texttt{WikiTable}.
So the approach of manually going
over high-quality \texttt{WikiTable} to curate mapping relationships
is unlikely to be sufficient.


The \texttt{WebTable} approach uses raw
tables similar to \texttt{WikiTable},
but considers tables not limited to the Wikipedia domain and thus has substantially 
better recall. While the precision of \texttt{WebTable}
and \texttt{Synthesis} are comparable, the recall of \texttt{Synthesis}
is substantially higher (0.88 vs. 0.32). Despite this, 
we want to note that the setup of this comparison of
is very favorable for \texttt{WebTable} --
we select the best table among the hundreds of millions of raw tables in \texttt{WebTable},
whereas in \texttt{Synthesis} we only use 
relations synthesized from over 8 website 
domains that is three orders of magnitude less (Section~\ref{sec:curation}).
Because it is not possible
for human to go over millions of tables to pick
useful mappings in practice, \texttt{WebTable} only provides an upper-bound
of what can be achieved and not really a realistic solution.

\texttt{UnionDomain} and \texttt{UnionWeb} 
synthesize tables based on table column names and 
domain names. The recall of these two approaches is 
considerably better than \texttt{WikiTable} and \texttt{WebTable},
showing the benefit of performing table synthesis.
However, this group of approaches merge tables 
only based on column names,
which are known to be uninformative 
and undescriptive in many cases. We observe that when applied
to the whole web, this often leads to over-grouping
and under-grouping. The overall f-scores of these
approaches are the best among all existing methods, but
still lag behind \texttt{Synthesis}, which uses values
that are more indicative of table compatibility.

\texttt{SynthesisPos} uses the same algorithm as
\texttt{Synthesis} but does not consider the
negative incompatibility induced by FDs. 
It is interesting to observe that result quality suffers 
substantially, which underlines the importance of the
negative signals.

\texttt{SchemaCC} performs substantially
worse than \texttt{Synthesis}. Recall that it uses 
the same positive/negative signals, but aggregate
pair-wise match decisions using connected components.
This simple aggregation tends to over-group and under-group
different tables, producing undesirable table clusters.

\texttt{SchemaPosCC} ignores the negative
signals used in \texttt{SchemaCC}, since FD-induced
negative signals are not explored in schema matching.
Unsurprisingly, result quality drops even further.

\texttt{Correlation} is similar to \texttt{SchemaCC}
that also mimics schema matchers with same signals, but
aggregate using correlation clustering. Overall, its f-score
is better than \texttt{SchemaCC}, but is still worse than
\texttt{Synthesis}. 
We think there are two main reasons why it does not work well. First, at the conceptual level, the objective of correlation clustering is the sum of positive and negative edges. Because the number of table pairs that would be in different clusters far exceeds the ones that should be in the same clusters, making negative edges dominate the objective function. However, in our problem, we should actually only care about whether tables in the same clusters correspond to the identical mapping, which are the intra-cluster positive edges that are more precisely modeled in our objective function. 
Second, a shortcoming of the parallel-pivot algorithm~\cite{Chierichetti14} is that it only looks at a small neighborhood for clusters (i.e. one-hop neighbors of cluster centers) for efficiency. When small tables in the same mapping form a chain of connected components, looking at the immediate neighborhood of a pivot (cluster center) will misses most other tables, producing results with low recall.

We implemented the collective schema-matching
method \texttt{WiseIntegrator}. It performs reasonably well
but still lags behind \texttt{Synthesis}, mainly because of the
difference in how scores are aggregated to produce holistic matches.

Please see Appendix~\ref{sec:moreexp} and~\ref{sec:moreAnalysis} for more analysis on the experimental results.

%
%
%

\subsection{Run-time Comparison}

We analyze the the complexity of our approach in this section. The basic input of our problem is a graph $G=(V, E)$ where $V$ represents candidate tables and $E$ represents their similarity. 
The most expensive part of our algorithm is in table synthesis (Step 2) that computes edge
similarity and performs iterative grouping. 

\begin{figure}\centering
   \begin{minipage}{0.28\textwidth}
      \includegraphics[width=1.\columnwidth]{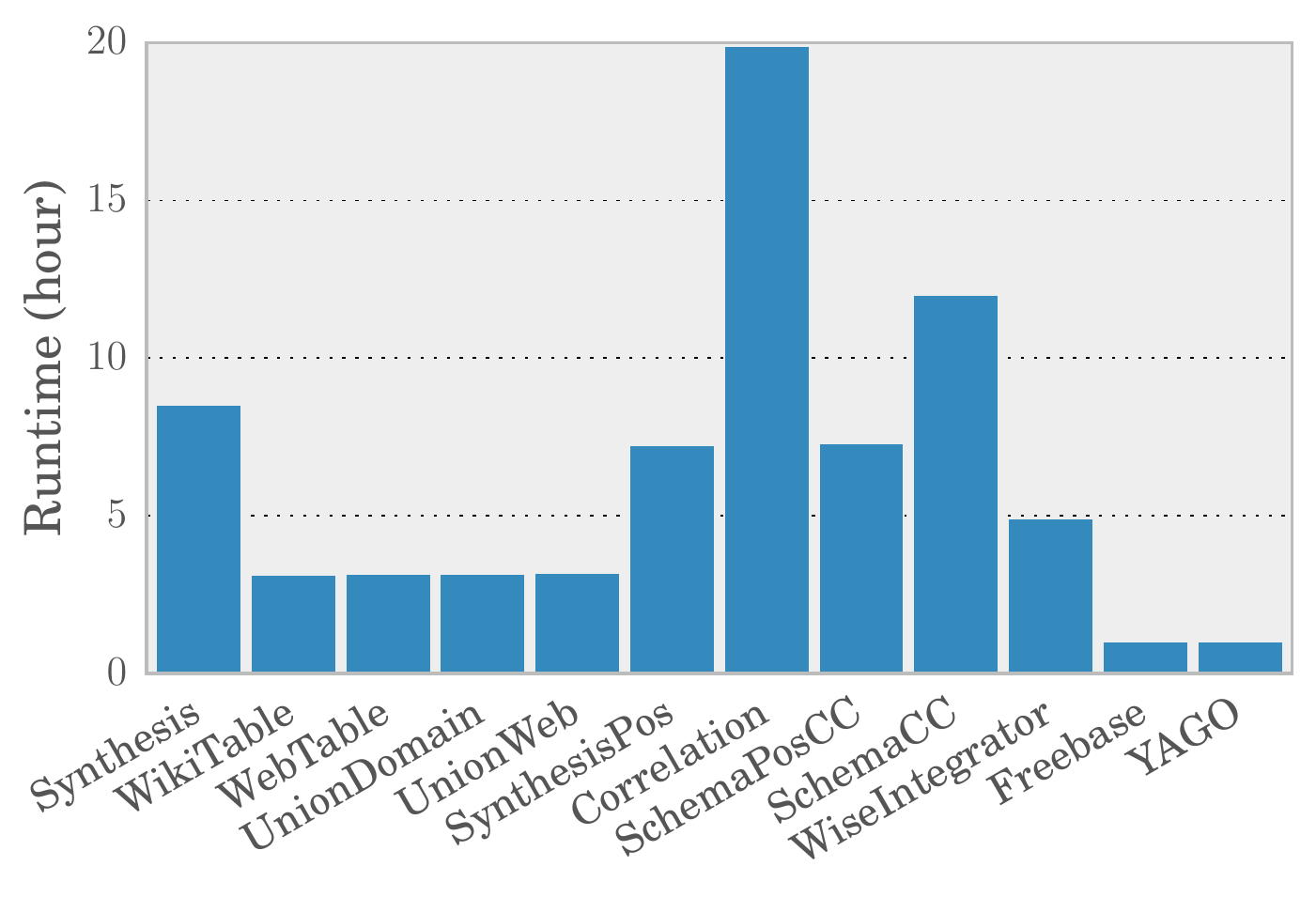}
      \caption{Runtime.}\label{fig:runtime}
   \end{minipage}
   \hfill 
   \begin{minipage}{0.18\textwidth}
     \includegraphics[width=\linewidth]{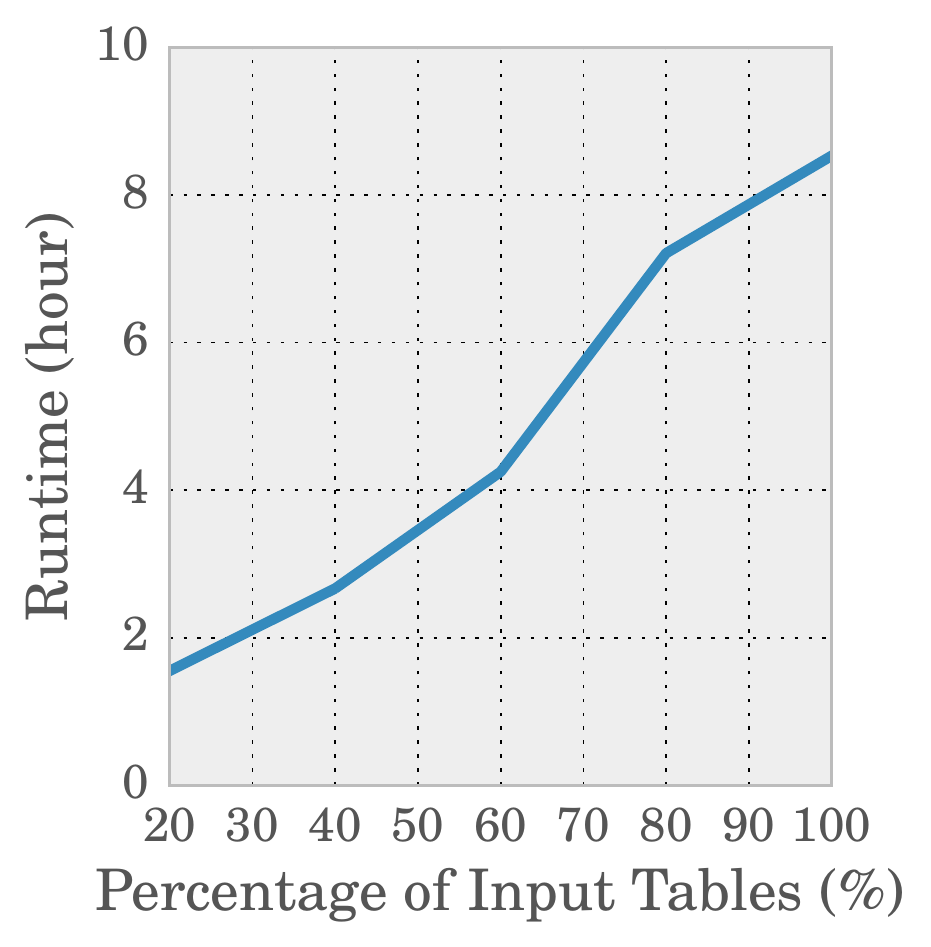}
     \caption{Scalability.}\label{fig:scale}
   \end{minipage}
\end{figure}

Figure~\ref{fig:runtime} compares the runtime of all approaches. Knowledge bases are the most efficient because it amounts to a lookup of the relation with the highest f-score among all relations. \texttt{WikiTable}, \texttt{WebTable}, \texttt{UnionDomain}, \texttt{UnionWeb}, and \texttt{WiseIntegrator} are all relatively efficient but requires scans of large table corpus.
Our approach \texttt{Synthesis} usually finishes within 10 hours (e.g., using parameters suggested here). \texttt{Correlation} is clearly the slowest, as correlation clustering converges very slowly even using the state-of-the-art parallel implementation on map-reduce~\cite{Chierichetti14}.


To test scalability of the proposed method, we 
sample $\{20\%, 40\%, 60\%, 80\%\}$ of the input data and measures execution time, as shown in Figure~\ref{fig:scale}. The complexity of the algorithm depends on the number of edges $|E|$. In the worst case $|E|$ can be quadratic to the number of tables $|V|$, but in practice $|E|$ is usually almost linear to $|V|$ due to edge sparsity. Figure~\ref{fig:scale}
suggests that the algorithm scales close to linearly to the input data, which is encouraging as it should also scale to even larger data sets with billions of tables.



\subsection{Sensitivity Analysis}
We analyze the effect of parameters used in \texttt{Synthesis}.

\begin{itemize}[noitemsep, nolistsep, leftmargin=*, wide=0\parindent]
\item{$\theta$.} We use $\theta$ as a parameter when defining approximate mapping relationship, which is empirically set as $95\%$. When we vary $\theta$ between $93\%$ and $97\%$, the number of resulting mappings change very little (by up to $1\%$). We have also reverse-engineered by calculating the degree of approximation in desirable ground truth mappings
(e.g. \texttt{Springfield} $\rightarrow$ \texttt{Illinois} and \texttt{Springfield} $\rightarrow$ \texttt{Texas} will create a violation). $95\%$ is sufficient to ensure that desired mappings will not be pruned incorrectly in almost all cases.

\item{$\tau$.} This parameter controls when we determine two candidates conflict. 
Our results suggest that the quality is generally insensitive to small $\tau$. The performance peaks at around $-0.05$. In our other experiments we actually used $\tau = -0.2$ that
also produces good quality.

\item{$\theta_{overlap}$} is a parameter for efficiency that 
determines the number of pruned edges $|E|$ in 
our graph. As $\theta_{overlap}$ increases, $|E|$ drops quickly. 
The quality of resulting clusters are insensitive to $\theta_{overlap}$.

\item{$\theta_{edge}$.} We make $\theta_{edge}$ the threshold to filter out edges with insignificant positive weight. Our experiment suggests that $\theta_{edge}=0.85$ has the best
performance.

\end{itemize}


\subsection{Experiments on \large{\ent}}

As we discussed earlier, unlike the \web{} domain
where a large fraction of ground truth mappings can 
be constructed using common sense knowledge
and online data sources, the ground truth mappings in the \ent{}
domain is difficult to build.
We are not familiar with many enterprise-specific data
values and encodings in this corpus, which makes
ensuring completeness and correctness of these mappings
difficult. 

We build $30$ benchmark cases with best effort to ensure
completeness (for some mappings the ground truth may be in
master databases we have no access to). 
To put the quality numbers in perspective, we 
compare \texttt{Synthesis} with single-table based \texttt{EntTable}, 
which is similar to \texttt{WebTable} in \web{}. 
As Figure~\ref{ex:enterprise_eval} suggests that \texttt{Synthesis} 
achieve significantly higher recall by merging small tables. Its precision
is also high by avoiding merging conflicting content.

\begin{figure}
\small
\centering
\begin{scriptsize}
\begin{tabular}{|l|c|c|}
\hline
& \texttt{Synthesis} & \texttt{EntTable} \\ \hline
Avg. (F-score, Prc., Rcl.)  & (0.96, 0.96, 0.97) & (0.84, 0.99, 0.79)  \\ \hline
\end{tabular}
\end{scriptsize}
\caption{Comparison with the alternative on \ent{}}
\label{ex:enterprise_eval}
\end{figure}

Figure~\ref{ex:enterprise} shows some examples 
of mapping relationships produced. 
A large fraction of relationships are indeed important
mappings, such as
(\texttt{product-family} $\rightarrow$ \texttt{code}),
(\texttt{profit-center} $\rightarrow$ \texttt{code}),
(\texttt{data-center} $\rightarrow$ \texttt{region}), etc.
Most of these results are well-structured and look
consistent (shown in
the right column of Figure~\ref{ex:enterprise}),
which is a good indication that results produced are
of high quality.

Just like in the \web{} domain, applications
equipped with these mapping relationships and
some human curation can perform intelligent 
operations such as auto-join as discussed earlier.
We note that these mappings are specific
to this enterprise in question. Using tables
to build such relationships would be the 
only reasonable choice, since alternatives like
knowledge bases would not exist in enterprise
domains.

Inspecting the results produced in \ent{} does
reveal interesting issues. For example,
we observe that for certain mapping relationships, the results
are of low quality with mixed data values
and meta-data values (e.g., column headers). It turns out that
in spreadsheets, tables with complex structures
such as pivot tables are popular. These complex tables 
are usually not flat relational tables that create 
difficulty for correct extraction. 

Overall, given that rich mappings are produced 
for a completely different ~\ent{} corpus, 
we believe that this exercise shows the promise of 
the \texttt{Synthesis} approach to generalize and
produce mappings by just using a corpus of tables as input.

\subsection{Effect of Conflict Resolution}

Conflict resolution
improves the f-score for 48 out of the 80 cases tested. 
On average, the precision increases from $0.903$ to 
$0.965$, while the average recall only dips slightly 
from $0.885$ to $0.878$. Such improvements 
shows that this post-processing step is useful
in removing inconsistent value pairs without
affecting coverage.

Cases such as (\texttt{state} $\rightarrow$ \texttt{capital}) 
see the biggest improvement. These
relationships tend be confused with other relationships
that disagree only on a small number of values. 
For example, the relationship (\texttt{state} $\rightarrow$ \texttt{capital}) 
tends to be confused with 
 (\texttt{state} $\rightarrow$ \texttt{largest-city}) with only minor
disagreements such as \texttt{Washington}
and \texttt{Olympia} vs. \texttt{Washington}
and \texttt{Seattle}. These conflicting value pairs will get
mixed into results because for some subset of values there may 
not be sufficient incompatibility to prevent
merges from happening. The conflict resolution step 
helps to prune away such incorrect values.

We compare our conflict resolution with majority voting. The proposed approach has a slightly higher f-score than majority voting.
 Appendix~\ref{sec:moreAnalysis} has detailed results comparing the f-scores.

\begin{figure}[t]
\small
\centering
\begin{scriptsize}
\begin{tabular}{|c|c|}
\hline
Mapping Relationship & Example Instances \\ \hline
\multirow{2}{*}{(product-family, code) }  & (Access, ACCES), \\ & (Consumer Productivity, CORPO), ...  \\ \hline
\multirow{2}{*}{(profit-center, code) }  & (P10018, EQ-RU - Partner Support), \\ & (P10021, EQ-NA - PFE CPM), ...  \\ \hline
\multirow{2}{*}{(industry, vertical) }  & (Accommodation, Hospitality), \\ & (Accounting, Professional Services), ...  \\ \hline
\multirow{2}{*}{(ATU, country) }  & (Australia.01.EPG, Australia), \\ & (Australia.02.Commercial, Australia), ... \\ \hline
\multirow{2}{*}{(data-center, region) }  & (Singapore IDC, APAC), \\ & (Dublin IDC3, EMEA), ...  \\ \hline
\end{tabular}
\end{scriptsize}
\caption{Example mapping relationships and values, from the enterprise spreadsheets corpus}
\label{ex:enterprise}
\end{figure}

\vspace{1mm}


\section{Related Work} \label{sec:related}

Ling and Halevy et al. studied the problem of
stitching together web tables from the same domain
based on column names~\cite{Ling:2013}. 
When adapting this technique to generate mapping 
relationships for the whole Web, however, it tends to
lead over-grouping and low-quality mappings
(as we show in the experiments), because
column names are often undescriptive and 
too generic to be indicative of the true meanings~\cite{Cortez15}
(e.g., column names like \texttt{code} and \texttt{name} are common).

Knowledge bases such as Freebase~\cite{Bollacker:2008}, 
and YAGO~\cite{Suchanek:2007} curate important 
entity-relationships, some of which may be mapping relationships. However, 
the coverage of knowledge bases is low as they often miss important
mappings. For instance, YAGO has none of the example mappings listed
in Table~\ref{tab:examples}, 
while Freebase misses two (stock and airport). For mappings that
do exist in knowledge bases, there are typically no or very few
synonyms such as ones listed in Table~\ref{tab:synonym}.
Lastly, knowledge bases are expensive to build, yet their mappings only
cover the public Web domain, and does not generalize to
other domains such as enterprises.

There is a long and fruitful line of research on schema 
matching~\cite{Rahm01} that can 
suggest semantic correspondence between columns
for human users.
These matching relationships provide useful information
about positive compatibility 
between tables. However, using only positive signals of compatibility are
insufficient for an unsupervised algorithm 
to synthesize diverse tables on the web, since distinct
relationships can share substantial value overlap. 
We introduce \textit{negative incompatibility} 
specific to functional dependency observed by mapping relationships,
which is shown in experiments to be critical
for high-quality synthesis.

A notable branch in schema matching~\cite{Bronzi:2013, HeBin03, He03, Su06, Zhang:2011} deals with schemas extracted from Web forms collectively for matches. These techniques mainly use linguistic similarity of attribute names and distributions.
However, the input schemas are required to be homogeneous and from the same conceptual domain (e.g., all forms are required to be about books, or automobiles, but not mixed). Methods in this class are the closest to our problem in the schema matching literature -- we experimentally compare with a representative method from this class~\cite{He03}.

\iftoggle{fullversion}
{
Compared to the traditional schema matching, 
there are two key aspects that differentiate our work from existing 
schema matching. (1) Traditional schema matching is studied in the context of a 
small number of database schemas. In our problem, while we also ``match'' 
semantically compatible table columns, we have to deal with millions of schemas 
(223M for the Web data set), which is many orders of magnitude larger than 
previously studied. (2) As a consequence of the scale, we can no longer afford 
to ask humans to verify results produced by the traditional pair-wise ``match'' 
operator (\cite{Bernstein11, Rahm01}), which is designed to be recall-oriented 
with false-positives that human users are supposed manually filter out. Because 
in our problem pair-wise manual verification for millions of schemas is no longer feasible, 
we choose to group all compatible schemas and have them verified only at the group level,
which would be easier and more efficient for human curation.

Although our problem would appear more difficult than schema 
matching, it is still tractable because we are 
interested in a very specific type of schemas, that are two-column tables 
satisfying functional-dependencies. This induces strong constraints for schema compatibility (the negative signal we exploit), which has not been explored in the classical schema matching for general tables (the existing literature mostly uses single-column type information to infer incompatibility). Furthermore, by looking at schemas holistically instead of one-pair-at-a-time, it allows us to reason globally and actually produce better matches (e.g., if table B is mostly contained by table A, and table C is also contained by A, then even if B and C share little overlap, we may still be able to group B and C using these information holistically, which may not be possible for pair-at-a-time matching).
}
{}

Techniques such as the novel DataXFormer~\cite{Abedjan:2015} represent an alternative 
class of approaches that ``searches'' tables based on user input and asks users to select relevant results to fill/join. While this is already a great improvement, our experience 
suggests that in many cases the need to search, retrieve, read, and manually piece 
together results from multiple tables is too cumbersome for this to be a viable feature
in Google Doc or Microsoft Excel, where most users may not have the necessary 
experience to go through the full process. Like knowledge-bases used by search engines
today, we hope curating knowledge of mappings can make them easily accessible to
a large number of spreadsheet users.

Additional related work can be found in Appendix~\ref{sec:additional-related}.

\vspace{1mm}


\section{Conclusions and Future Work} \label{sec:conclusion}


In this paper, we study the problem of 
synthesizing mapping relationships using tables.
Our work is a first step in the direction
to facilitate the curation of mapping relationships.
Questions that we would like to address in the future
include: (1) how to best present related 
result clusters with overlapping values 
to human users to solicit feedback, so that users will not
be confused by clusters with repeating values;
(2) how to complement the corpus-driven approach
to better cover mappings with large numbers of instances,
by using other sources such as authoritative third-party data sets.
We hope our work will serve as a springboard for 
future research on the important problem of curating mapping
relationships.

\balance
{\scriptsize
\bibliographystyle{abbrv}
{\bibliography{reference}}

\begin{thebibliography}{10}

\bibitem{googlewebtables}
{Google Web Tables}.
\newblock \url{http://research.google.com/tables}.

\bibitem{powerbi}
{Microsoft Excel Power Query}.
\newblock \url{http://office.microsoft.com/powerbi}.

\bibitem{Huhtala99}
Tane: An efficient algorithm for discovering functional and approximate
  dependencies.
\newblock In {\em Computer Journal}, 1999.

\bibitem{Abedjan:2015}
Z.~Abedjan, J.~Morcos, M.~N. Gubanov, I.~F. Ilyas, M.~Stonebraker, P.~Papotti,
  and M.~Ouzzani.
\newblock Dataxformer: Leveraging the web for semantic transformations.
\newblock In {\em CIDR}, 2015.

\bibitem{Bejerano06}
Y.~Bejerano, M.~A. Smith, J.~Naor, and N.~Immorlica.
\newblock Efficient location area planning for personal communication systems.
\newblock In {\em Transaction of Networking}, 2006.

\bibitem{Bernstein11}
P.~A. Bernstein, J.~Madhavan, and E.~Rahm.
\newblock Generic schema matching, ten years later.
\newblock In {\em Proceedings of VLDB}, 2011.

\bibitem{Bollacker:2008}
K.~Bollacker, C.~Evans, P.~Paritosh, T.~Sturge, and J.~Taylor.
\newblock Freebase: A collaboratively created graph database for structuring
  human knowledge.
\newblock In {\em SIGMOD}, pages 1247--1250, 2008.

\bibitem{Broder:1997}
A.~Broder.
\newblock On the resemblance and containment of documents.
\newblock In {\em SEQUENCES}, pages 21--. IEEE Computer Society, 1997.

\bibitem{Bronzi:2013}
M.~Bronzi, V.~Crescenzi, P.~Merialdo, and P.~Papotti.
\newblock Extraction and integration of partially overlapping web sources.
\newblock {\em PVLDB}, pages 805--816, 2013.

\bibitem{synonyms}
K.~Chakrabarti, S.~Chaudhuri, Z.~Chen, K.~Ganjam, and Y.~He.
\newblock Data services leveraging bing's data assets.
\newblock {\em {IEEE} Data Eng. Bull.}, 2016.

\bibitem{Chen16}
Y.~Chen, S.~Goldberg, D.~Z. Wang, and S.~S. Johri.
\newblock Ontological pathfinding: Mining first-order knowledge from large
  knowledge bases.
\newblock In {\em SIGMOD}, 2016.

\bibitem{Chierichetti14}
F.~Chierichetti, N.~Dalvi, and R.~Kumar.
\newblock Correlation clustering in mapreduce.
\newblock In {\em KDD}, 2014.

\bibitem{Chitnis:2013}
L.~Chitnis, A.~Das~Sarma, A.~Machanavajjhala, and V.~Rastogi.
\newblock Finding connected components in map-reduce in logarithmic rounds.
\newblock In {\em ICDE}, pages 50--61, 2013.

\bibitem{Church:1990}
K.~W. Church and P.~Hanks.
\newblock Word association norms, mutual information, and lexicography.
\newblock {\em Comput. Linguist.}, 16(1):22--29, 1990.

\bibitem{Cortez15}
E.~Cortez, P.~A. Bernstein, Y.~He, and L.~Novik.
\newblock Annotating database schemas to help enterprise search.
\newblock {\em Proceedings of VLDB}, 2015.

\bibitem{Dahlhaus:1994}
E.~Dahlhaus, D.~S. Johnson, C.~H. Papadimitriou, P.~D. Seymour, and
  M.~Yannakakis.
\newblock The complexity of multiterminal cuts.
\newblock {\em SIAM J. Comput.}, 23(4):864--894, 1994.

\bibitem{Demaine:2006}
E.~D. Demaine, D.~Emanuel, A.~Fiat, and N.~Immorlica.
\newblock Correlation clustering in general weighted graphs.
\newblock {\em Theor. Comput. Sci.}, 361(2):172--187, 2006.

\bibitem{Galarraga13}
L.~A. Gal\'{a}rraga, C.~Teflioudi, K.~Hose, and F.~Suchanek.
\newblock Amie: Association rule mining under incomplete evidence in
  ontological knowledge bases.
\newblock In {\em WWW}, 2013.

\bibitem{garg1994multiway}
N.~Garg, V.~V. Vazirani, and M.~Yannakakis.
\newblock Multiway cuts in directed and node weighted graphs.
\newblock In {\em ICALP}, pages 487--498, 1994.

\bibitem{Gupta:2014}
R.~Gupta, A.~Halevy, X.~Wang, S.~E. Whang, and F.~Wu.
\newblock Biperpedia: An ontology for search applications.
\newblock {\em PVLDB}, pages 505--516, 2014.

\bibitem{HeBin03}
B.~He and K.~C.-C. Chang.
\newblock Statistical schema matching across web query interfaces.
\newblock In {\em Proceedings of SIGMOD}, 2003.

\bibitem{He04}
H.~He, W.~Meng, C.~Yu, and Z.~Wu.
\newblock Wise-integrator: An automatic integrator of web search interfaces for
  e-commerce.
\newblock {\em VLDB Journal}, 2004.

\bibitem{He03}
H.~He, W.~Meng, C.~T. Yu, and Z.~Wu.
\newblock Wise-integrator: An automatic integrator of web search interfaces for
  e-commerce.
\newblock In {\em PVLDB}, 2003.

\bibitem{He15}
Y.~He, K.~Ganjam, and X.~Chu.
\newblock Sema-join: joining semantically-related tables using big table
  corpora.
\newblock In {\em Proceedings of VLDB}, 2015.

\bibitem{Hopcroft:1973}
J.~E. Hopcroft and J.~D. Ullman.
\newblock Set merging algorithms.
\newblock {\em SIAM Journal on Computing}, 2(4):294--303, 1973.

\bibitem{Hu:1963}
T.~C. Hu.
\newblock Multi-commodity network flows.
\newblock {\em Operations Research}, 11(3):344--360, 1963.

\bibitem{Kimball02}
R.~Kimball and M.~Ross.
\newblock {\em The Data Warehouse Toolkit: The Definitive Guide to Dimensional
  Modeling}.
\newblock 2002.

\bibitem{Li16}
Y.~Li, J.~Gao, C.~Meng, Q.~Li, L.~Su, B.~Zhao, W.~Fan, and J.~Han.
\newblock A survey on truth discovery.
\newblock In {\em SIGKDD Exploration}, 2016.

\bibitem{Lin10}
T.~Lin, Mausam, and O.~Etzioni.
\newblock Identifying functional relations in web text.
\newblock In {\em Proceedings of EMNLP}, 2010.

\bibitem{Ling:2013}
X.~Ling, A.~Halevy, F.~Wu, and C.~Yu.
\newblock Synthesizing union tables from the web.
\newblock In {\em IJCAI}, pages 2677--2683, 2013.

\bibitem{Rahm01}
E.~Rahm and P.~A. Bernstein.
\newblock A survey of approaches to automatic schema matching.
\newblock {\em The VLDB Journal}, 10(4):334--350, Dec. 2001.

\bibitem{Ritter08}
A.~Ritter, D.~Downey, S.~Soderland, and O.~Etzioni.
\newblock It's a contradiction---no, it's not: A case study using functional
  relations.
\newblock In {\em EMNLP}, pages 11--20, 2008.

\bibitem{Su06}
W.~Su, J.~Wang, and F.~Lochovsky.
\newblock Holistic schema matching for web query interfaces.
\newblock In {\em Proceedings of EDBT}, 2006.

\bibitem{Suchanek:2007}
F.~M. Suchanek, G.~Kasneci, and G.~Weikum.
\newblock Yago: A core of semantic knowledge.
\newblock In {\em WWW}, pages 697--706, 2007.

\bibitem{Ukkonen:1985}
E.~Ukkonen.
\newblock Algorithms for approximate string matching.
\newblock {\em Inf. Control}, 64(1-3):100--118, 1985.

\bibitem{varizani}
V.~Varizani.
\newblock {\em Approximation algorithms}.
\newblock Springer Verlag, 2001.

\bibitem{full}
Y.~Wang and Y.~He.
\newblock Synthesizing mapping relationships using table corpus.
  \url{https://www.microsoft.com/en-us/research/wp-content/uploads/2017/03/mapping-synthesis-full.pdf}.
\newblock Technical report.

\bibitem{Yakout12}
M.~Yakout, K.~Ganjam, K.~Chakrabarti, and S.~Chaudhuri.
\newblock Infogather: Entity augmentation and attribute discovery by holistic
  matching with web tables.
\newblock In {\em SIGMOD}, 2012.

\bibitem{Yannakakis:1983}
M.~Yannakakis, P.~C. Kanellakis, S.~S. Cosmadakis, and C.~H. Papadimitriou.
\newblock Cutting and partitioning a graph after a fixed pattern.
\newblock In J.~Diaz, editor, {\em ICALP}, pages 712--722, 1983.

\bibitem{Zhang:2011}
M.~Zhang, M.~Hadjieleftheriou, B.~C. Ooi, C.~M. Procopiuc, and D.~Srivastava.
\newblock Automatic discovery of attributes in relational databases.
\newblock In {\em SIGMOD}, pages 109--120, 2011.

\end{thebibliography}
}

\nocite{full}

\vspace{-0.2cm}
\appendix


\section{Table Extraction}
\label{apx:filter}

\begin{algorithm}[!h]
{\scriptsize
\caption{Candiate Extraction} \label{alg:candidateExtraction}
\DontPrintSemicolon
\KwIn{Table corpus $\TableCorpus$}
\KwOut{Candidate two-column table set $\TwoColTblCorpus$}

$\TwoColTblCorpus \Leftarrow \emptyset$\;
\ForEach{$T \in \TableCorpus$} {
    $T' \Leftarrow \emptyset$\;
    \ForEach {$C_i \in T$} {
        \If {$C_i$ is \emph{not} removed by PMI filter} {
            $T' \Leftarrow T' \cup \{C_i\}$\;
        }
    }
    \ForEach {$C_i, C_j \in T'~~(i \neq j)$} {
        $B \Leftarrow (C_i, C_j)$\;
        \If {$B$ is \emph{not} removed by FD filter} {
            $\TwoColTblCorpus \Leftarrow \TwoColTblCorpus \cup \{B\}$\;
        }
    }
}
}
\end{algorithm}
\vspace{-1mm}

Algorithm~\ref{alg:candidateExtraction} gives the 
pseudo-code for candidate table extraction. The two
steps correspond to PMI-based filtering and
FD-based filtering, respectively.

\section{Approximate String Matching}
\label{apx:approx-string}

\begin{algorithm}
{\scriptsize
\DontPrintSemicolon
\caption{Approximate String Matching} \label{alg:approximateMatching}
\KwIn{Strings $v_1$ and $v_2$, distance bound $\theta_{ed}$}
\KwOut{Boolean $Matched$}

\If{$|v_1| > |v_2|$} {
    swap($v_1, v_2$)\;
}
$dist_{|v_1|, |v_2|} \Leftarrow \infty$\;
$dist_{i, 0} \Leftarrow i,~~~~\forall (i \in 0..|v_1|)$\;
$dist_{0, j} \Leftarrow j,~~~\forall (j \in 0..|v_2|)$\;
\For{$i \in 1..|v_1|$} {
    $lower \Leftarrow \max\{1, i - \theta_{edit}\}$\;
    $upper \Leftarrow \min\{|v_2|, i + \theta_{edit}\}$\;
    \For {$j \in lower..upper$} {
        $dist_{i,j} \Leftarrow \infty$\;
        \If{$dist_{i-1,j} \neq$ NULL}{$dist_{i,j} \Leftarrow \min\{dist_{i-1,j} + 1, dist_{i,j}\}$}
        \If{$dist_{i,j-1} \neq$ NULL}{$dist_{i,j} \Leftarrow \min\{dist_{i,j-1} + 1, dist_{i,j}\}$}
        \If{$dist_{i-1,j-1} \neq$ NULL}{$dist_{i,j} \Leftarrow \min\{dist_{i-1,j-1} + \mathds{1}\{v_{1}[i] \neq v_{2}[j]\}, dist_{i,j}\}$}
    }
}
$Matched \Leftarrow (dist_{|v_1|, |v_2|} \leq \theta_{ed})$\;
}
\end{algorithm}

The algorithm for efficient approximate string matching
is shown in Algorithm~\ref{alg:approximateMatching}.
We leverage the fact that the desired distance $\theta_{ed}$
is often small to only perform dynamic programming on a narrow band 
in the diagonal direction of the matrix instead of performing a full DP, which
is in spirit similar to 
Ukkonen's algorithm~\cite{Ukkonen:1985}.

\section{Proof of Theorem~\ref{1}}
\label{apx:hardness}
\begin{proof}
We prove it by showing that Problem~\ref{problem:partitioning} is a more general case of a typical multi-cut problem in a weighted graph~\cite{Hu:1963}. 
Given an undirected graph $G_C=(V_C, E_C)$, a weight function $w_C$ of the edges, and a set of $k_C$ pairs of distinct vertices $(s_i, t_i)$, the multi-cut problem is to find the minimum weight set of edges of $G_C$ that disconnect every $s_i$ from $t_i$. The multi-cut problem is NP-hard~\cite{Dahlhaus:1994}.

Now we transform $G_C=(V_C, E_C)$ to graph $G=(\TwoColTblCorpus, E)$ as follows: (i) We first divide the weights by a large number, $\max\{w_C(v_i,v_j)\}$, to change the range of weights to $(0, 1]$. (ii) We define $\TwoColTblCorpus=V_C$ and $E=E_C$. (iii) We make positive weights $w^+(v_i, v_j)=w^+(v_j, v_i)=w_C(v_i, v_j)$. (iv) For each pair of vertices $(s_i, t_i)$, we make negative weights $w^-(s_i, t_i)=w^-(t_i, s_i)=-1<\tau$.

As a result, each partitioning $\PartitionSet$ in Problem~\ref{problem:partitioning} corresponds to exactly one cut $E_{cut}$ in the above multi-cut problem because: (i) Constraint~\eqref{eq:conflict} guarantees that $s_i$ and $t_i$ are never in the same partition. (ii) The edges across partitions are $E_{cut}$ (i.e. the set of edges to be removed) in the multi-cut problem. (iii) Let
$w^+(\PartitionSet)$ be the objective function $\sum_{P \in \PartitionSet}w^+(P)$ of Problem~\ref{problem:partitioning},
the sum of weights of the graph be $w_C(G_C)$, and the weight of cut be $w_C(E_{cut})$.
Then $w^+(\PartitionSet)+w_C(E_{cut})=w_C(G_C)$. So maximizing $w^+(\PartitionSet)$ is equivalent to minimizing $w_C(E_{cut})$. Therefore we reduce the multi-cut problem to Problem~\ref{problem:partitioning}.

So Problem~\ref{problem:partitioning} is NP-hard.
\end{proof}

\iftoggle{fullversion}
{
\section{LP Relaxation \& Approximation}
\label{apx:lp}

Problem~\ref{problem:ILP} is formulated as an integer linear program.
We can relax it by replacing the integrality constraint
with  $d_{ij} \in [0, 1]$ to make it an LP, which
can then be solved using a standard solver
in polynomial time. 

Using the optimal fractional solution from the LP,
one can round such a solution in a 
region-growing procedure~\cite{Bejerano06, garg1994multiway, varizani} 
that finds an integral solution close to the fractional
solution. This randomized rounding process guarantees
$O(\log N)$ approximation for the loss minimization
version of the problem.

As discussed earlier, the number of variables $d_{ij}$ required in this
approach is quadratic to the number of candidate tables. 
Given millions of candidates and a quadratic number of variables,
in practice solving the resulting LP is difficult. So despite
the nice theoretical guarantees, we are not using this
LP-based rounding approach due to the scale required
in our problem.
}
{
}

\section{Table Synthesis by Partitioning}
\label{apx:table-partition-algo}

\begin{algorithm}[t]
{\scriptsize
\caption{Table-Synthesis by Partitioning} \label{alg:greedy}
\DontPrintSemicolon
\KwIn{Graph $G=(\mathcal{B}, E)$, threshold $\tau$}
\KwOut{Set of Partitions $\PartitionSet$}

$P(B_i) \Leftarrow \{B_i\},~~\forall B_i \in \mathcal{B}$\; \label{line:newGraphStart}
$\mathcal{B}_P \Leftarrow \bigcup_{B_i \in \mathcal{B}}\{P(B_i)\}$\;
$E_{P} \Leftarrow \bigcup_{(B_i, B_j) \in E}\{(P(B_i), P(B_j))\}$\;
$w^+_P(P(B_i), P(B_j)) \Leftarrow w^+(B_i, B_j)$\;
$w^-_P(P(B_i), P(B_j)) \Leftarrow w^-(B_i, B_j)$\;
$G_P \Leftarrow (\mathcal{B}_P, E_P)$\; \label{line:newGraphEnd}

\While{$true$} {
    $e(P_1, P_2) \Leftarrow \underset{P_1 \neq P_2, w^-_P(P_1, P_2)\geq \tau}{\argmax} (w^+_P(P_1, P_2))$\; \label{line:findSimilar}
    \If{$e=$NULL} {
        \textbf{break}\;
    }
    $P' \Leftarrow P_1 \cup P_2$\; \label{line:mergeStart}
    Add $P'$ and related edges into $\mathcal{B}_P$ and $E_P$\;
    \ForEach{$P_i \notin \{P_1, P_2\}$} {
        $w^+_P(P_i, P') \Leftarrow w^+_P(P', P_i) \Leftarrow w^+_P(P_i, P_1) + w^+_P(P_i, P_2)$\;
        $w^-_P(P_i, P') \Leftarrow w^-_P(P', P_i)  \Leftarrow \min \{w^-_P(P_i, P_1), w^-_P(P_i, P_2)\}$\;
    }
    Remove $P_1, P_2$ and related edges from $\mathcal{B}_P$ and $E_P$\; \label{line:mergeEnd}
}
$\PartitionSet \Leftarrow \mathcal{B}_P$\;
}
\end{algorithm}
Algorithm~\ref{alg:greedy} shows the pseudo-code for table synthesis.

\section{Scalability of Iterative Partition}
\label{apx:connected}


We use the Hash-to-Min algorithm to compute connected 
components on Map-Reduce~\cite{Chitnis:2013}. 
This algorithm treats every vertex and its neighbors as a cluster initially. 
Then for each cluster, it sends a message of the cluster ID to all its members.
Next every vertex chooses the minimum cluster ID it receives and propagate this minimum ID as the new ID of all the other clusters who sends message to it. The algorithm iteratively apply the above steps until convergence. This algorithm solves our problem very efficiently. 

Now given a subgraph, we apply Algorithm~\ref{alg:greedy} to solve Problem~\ref{problem:partitioning}. Set union and lookup are two frequent operations in Algorithm~\ref{alg:greedy}. So we use a disjoint-set data structure to speed 
up the process~\cite{Hopcroft:1973}. Its idea is to maintain a tree to represent each set so that union and lookup of the tree root are much faster than a na\"ive set operation.

\section{Proof of Hardness for Conflict Resolution}
\label{apx:truth}
\begin{proof}
We prove the hardness of conflict resolution by reducing the maximum independent set (MIS) problem to it. Given a graph $G_M=(V_M, E_M)$ in MIS, we correspondingly build a partition $P=\{B_1, B_2, ...\}$ in Problem~\ref{problem:truth} as follows: (1) For each $v_m \in V_M$, we create a $B_m$. (2) For each $e_m(v_i, v_j) \in E_M$, we create a pair of contradicting value pairs $((l, r), (l, r'))$. We add $(l, r)$ to $B_i$ and $(l, r')$ to $B_j$. (3) Let the maximum vertex degree of $G_M$ be $deg$. We add dummy value pairs to each $B_i$ to make $|B_i|=deg$. These dummy value pairs do not conflict with any existing value pairs. Obviously, the MIS problem has a solution with size $S_{MIS}$, if and only if Problem~\ref{problem:truth} has a solution with weight $S_{MIS} \cdot deg$.
\end{proof}

\begin{figure}[t]
\vspace{-1mm}
\centering
{\fontsize{6}{7.2} \selectfont 
\begin{tabular}{|l|l|}
\hline
Mapping Relationship & Example Instances \\ \hline \hline
\multirow{2}{*}{(US-city, state-abbr.)}  & (New York, NY), \\ & (Chicago, IL), ...  \\ \hline
(gun-powder-name,  & (Varget, Hodgdon), \\ 
company) & (RL-15, Alliant), ...  \\ \hline
(UK-county,   & (Suffolk, England), \\
country) & (Lothian, Scotland), ...  \\ \hline
(India-railway-station, & (Vadodara Junction, Gujarat), \\ 
state) & (Itarsi Junction, Madhya Pradesh), ...  \\ \hline
(wind, & (gentle breeze, 3), \\ 
Beaufort-scale) & (storm, 10), ...  \\ \hline
(state/province abbr., & (QLD, AU), \\ 
country) & (ON, CA), ...  \\ \hline
(ISO3166-1-Alpha-3, & (USA, US), \\ 
ISO3166-1-Alpha-2) & (FRA, FR), ...  \\ \hline
\multirow{2}{*}{(movie, year)} & (Pulp Fiction, 1994), \\ 
 & (Forrest Gump, 1994), ...  \\ \hline
\multirow{2}{*}{(movie, distributor)} & (The Dark Knight Rises, WB), \\ 
 & (Life of Pi, Fox), ...  \\ \hline
(ODBC-configuration, & (odbc.check\_persistent, on), \\ 
default-value) & (odbc.default\_db, no value), ...  \\ \hline
\multirow{2}{*}{(automobile, type)} & (F-150, truck), \\ 
 & (Escape, SUV), ...  \\ \hline
(family-member, & (Mother, F), \\ 
gender) & (Brother, M), ...  \\ \hline
\multirow{2}{*}{(ASCII-abbr., code)} & (NUL, 0), \\ 
 & (ACK, 6), ...  \\ \hline
(ISO-4217-currency-  & (USD, 840), \\ 
code, num) & (EUR, 978), ...  \\ \hline
\end{tabular}
}
\vspace{-0.25cm}
\caption{Additional mappings synthesized from \web{}.}
\label{ex:web}
\vspace{-1.5mm}
\end{figure}

\section{Conflict Resolution}
\label{apx:postprocessing}

\begin{algorithm}
{\scriptsize
\DontPrintSemicolon
\caption{Conflict Resolution} \label{alg:truthDiscovery}

\KwIn{Partition $P=\{B_1, B_2, ...\}$}
\KwOut{$P_T$ without conflict} 

$P_T \Leftarrow P$\;
\While{$\exists B_i, B_j \in P_T, |F(B_i, B_j)| > 0$} {
    $\textit{InstSet} \Leftarrow \bigcup_{B_i \in P_T} B_i$\; \label{line:truthInstSetBegin}
    \ForEach{$(v_{1}, v_{2}) \in \textit{InstSet}$} {
        $cnt_V(v_{1}, v_{2}) \Leftarrow \#~\text{conflicting value pairs in}~\textit{InstSet}$\;
    } \label{line:truthInstSetEnd}
    \ForEach{$B_i \in P_T$ \label{line:truthCandBegin}} {
        $cnt_B(B_i) \Leftarrow \max_{(v_1, v_2) \in B_i} \{cnt_V(v_1, v_2)\}$\;
    } 
    $B_i \Leftarrow \argmax_{B_i \in P_T} cnt_B(B_i)$\; 
    $P_T \Leftarrow P_T \textbackslash \{B_i\}$\; \label{line:truthCandEnd}
}
}
\end{algorithm}

Algorithm~\ref{alg:truthDiscovery} iteratively finds value pairs that 
conflict with the most other value pairs and removes its candidate table. 
Specifically, given a value pair $(v_1, v_2)$, Line~\ref{line:truthInstSetBegin} to Line~\ref{line:truthInstSetEnd} counts the number of conflicting value pairs. Line~\ref{line:truthCandBegin} to Line~\ref{line:truthCandEnd} 
finds the candidate that introduces the most conflicts and removes it. 
In practice, we maintain an index for each value pair and each candidate 
to keep track the number of conflicts. We use a heap that supports 
update to select the most conflicting candidate efficiently.

\section{Table Expansion} \label{apx:table-expansion}

We see that synthesized relationships provide a robust
``core'', which can be used to bring in additional instances.
We perform an optional expansion step, by using external
data resources such as data.gov or spreadsheet 
files (\emph{.xlsx}) crawled from other trustable web 
sources, that are more likely to be comprehensive
(web tables on the other hand are often for human consumption
and tend to be short). We compute 
the similarity and dissimilarity between our synthesized ``cores'' 
and these external sources, and merge if certain requirements are met. 
Note that this step can also happen at curation time with human 
users in the loop. 

\iftoggle{fullversion}
{We compare the f-score 
before and after table expansion. 
Overall the effect is limited. F-score is improved 
substantially for only two cases, namely
(\texttt{airport-name} $\rightarrow$ \texttt{IATA-code}) and 
(\texttt{airport-name} $\rightarrow$ \texttt{ICAO-code}).
These two have over 10k instances in
ground truth. So synthesis alone is not sufficient
to recover the full relationship, and expansion brings more
pronounced effect.
}
{}

\section{More Example Mappings}
\label{sec:moreexp}
Here we discuss additional synthesized mappings that are not in 
the benchmark.  
Figure~\ref{ex:web} lists additional popular 
mappings synthesized using \web. Many relationships involve geographic 
information such as (\texttt{US-city} $\rightarrow$ \texttt{state-abbreviation}), 
$\,$(\texttt{India-railway-station} $\rightarrow$ \texttt{state}), 
$\,$(\texttt{UK-}\texttt{county} $\rightarrow$ \texttt{country}), etc. 
There are a variety of other relationships 
such as (\texttt{wind} $\rightarrow$ \texttt{Beaufort-scale}), 
$\,$(\texttt{ASCII-abbre-viation} $\rightarrow$ \texttt{code}), 
(\texttt{automobile} $\rightarrow$ \texttt{type}) etc. 
We find reasonable meanings of these binary relationships and
consider them to be high quality.

However, certain synthesized binary relationships 
are less ideal as mappings. We show such cases
in Figure~\ref{ex:webBad}. For example, certain relationships 
are temporal that only hold for a period of time. Examples
like (\texttt{F1-driver}, \texttt{team}), (\texttt{English-football-club}, \texttt{points}), (\texttt{college-football-team}, \texttt{ranking}) are in this category.
Because this leads to many mappings of the same type 
that are true in different point of time (e.g., points of soccer teams),
additional reasoning of conflicts between synthesized clusters
can potentially identify such temporal mappings. We leave 
improving results in this regard as future work.

Certain tables are used repeatedly for formatting purpose, whose
values would get extracted as popular mappings. For example,
the (\texttt{month}, \texttt{month}) in Figure~\ref{ex:webBad} 
maps \texttt{January} to \texttt{July}, \texttt{Feb} to \texttt{August} 
and so on, simply because many pages list 12 month calendar 
as two column tables. Results in this category are not significant
in numbers, and should be relatively easy for human to prune out.


\begin{figure}[t]
\vspace{-1mm}
\small
\centering
{\fontsize{6}{7.2} \selectfont 
\begin{tabular}{|l|l|}
\hline
Mapping Relationship & Example Instances \\ \hline \hline
\multirow{2}{*}{(MiLB-leagues, level)} & (PCL, AAA), \\ 
 & (IL, AAA), ...  \\ \hline 
 \multirow{2}{*}{(baseball-team, league)} & (NYY, AL), \\ 
 & (LAD, NL), ...  \\ \hline
(English-football-club,  & (Manchester City, 16), \\ 
points) & (Liverpool, 17), ...  \\ \hline
(US-soccer-club, & (Houston Dynamo, 48), \\ 
points) & (Chicago Fire, 49), ...  \\ \hline 
\multirow{2}{*}{(F1-driver, team)} & (Sebastian Vettel, Ferrari), \\ 
 & (Lewis Hamilton, Mercedes), ...  \\ \hline 
(college-football-team,  & (Alabama, 1), \\ 
ranking) & (Clemson, 3), ...  \\ \hline
(college-football-team,  & (Stanford, 5-0), \\ 
score) & (Michigan, 5-0), ...  \\ \hline
(football-player,  & (Marques Colston, NO), \\ 
team) & (Victor Cruz, NYG), ...  \\ \hline
\multirow{2}{*}{(month, month)} & (January, July), \\ 
 & (February, August), ...  \\ \hline 
\multirow{2}{*}{(day, hour)} & (Monday, 7:30AM - 5:30PM), \\ 
 & (Tuesday, 7:30AM - 5:30PM), ...  \\ \hline 
%
\end{tabular}
}
\vspace{-0.25cm}
\captionof{figure}{Synthesized relationships not ideal as mappings.}
\label{ex:webBad}
\vspace{-1.5mm}
\end{figure}

\paragraph*{Usefulness of Mappings} We sample 
the top clusters produced based on popularity (the number of tables/domains
contributing to the cluster). 
We classify the mapping corresponding to each cluster into three categories: Meaningful mapping (static), Meaningful mapping (temporal), and Meaningless mapping.  
For top 500 clusters we 
inspected, 
$49.6\%$ are static, $37.8\%$ are temporal, and only $12.6\%$ are meaningless,
which is encouraging.


\section{More Experimental Results} \label{sec:moreAnalysis}

\begin{figure*}[!ht]
\centering
\includegraphics[width=.98\textwidth]{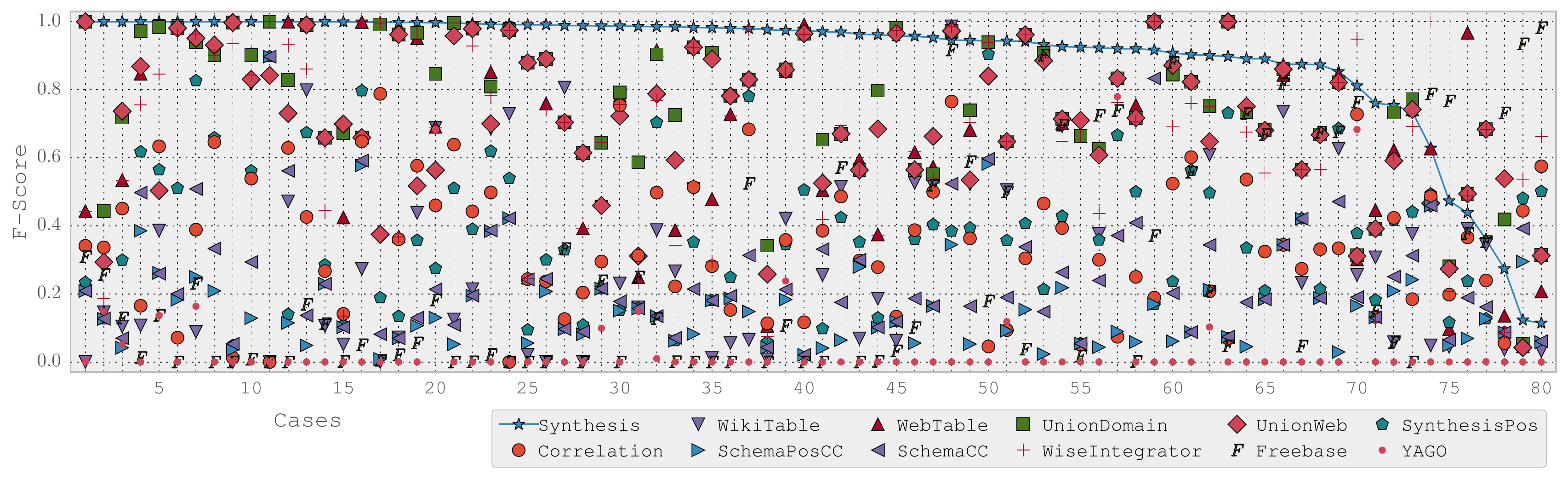}
\caption{Comparison with alternatives on individual cases (Sorted by f-score of our Synthesis approach).}\label{fig:baselineDetail}
\end{figure*}

Methods using knowledge bases \texttt{Freebase} and \texttt{YAGO} 
have reasonable precision, which is expected because 
they are extensively curated. The recall numbers of these
methods, however, are
substantially lower, because 
a significant fraction of useful mappings are
missing from existing knowledge bases. 
This indicates that knowledge bases alone
are unlikely to be insufficient for harvesting rich mappings.

Figure~\ref{fig:baselineDetail} gives detailed quality numbers 
for individual cases in the benchmark. The overall observation here is
consistent with Figure~\ref{fig:baseline}. We can see that
for a large fraction of test cases, \texttt{Synthesis} produces
results of high quality, which are amenable to further
human curation before they are applied in
data-driven applications. It is interesting to note that even
when high-precision methods \texttt{WikiTable} 
and \texttt{Freebase} already have a complete 
table covering instances in 
certain benchmark cases such as \texttt{chemical element} and
\texttt{country code}, their f-scores are still
low despite almost perfect precision. This is because
the ideal ground truth mapping should contain many synonymous
names for the same entity (e.g., shown in Figure~\ref{tab:synonym} for
\texttt{country code}). In fact, the results \texttt{Synthesis} produces
have over 470 entries for \texttt{country code} 
(compared to around 200 distinct countries), and 
over 200 entries for \texttt{chemical element} 
(compared to about 100 distinct ones). Methods
like \texttt{WikiTable} and \texttt{Freebase} tend
to have only one name mention for the same entity 
in one table, thus producing inferior scores for recall. 
As we have discussed in the introduction, such synonymous
entity names are important for
applications like auto-join and auto-correct, since
a user data table can always have one name but not the other.


Interestingly, for a number of cases
where \texttt{Synthesis} does not produce satisfactory
results (towards the right of the figure), 
\texttt{Freebase} performs surprisingly well.
It appears that for domains like chemicals,
mappings such as (Case 80: \texttt{chemical-compound} $\rightarrow$ \texttt{formula}) 
and (Case 74: \texttt{substance} $\rightarrow$ \texttt{CAS number}) have little
web presence, which gives limited scope for synthesis using tables.
On the other hand, \texttt{Freebase} has many structured
data sets curated by human from specialized data sources
covering different domains, thus providing better coverage
where no techniques using web tables gives reasonable performance.
We believe this shows that knowledge bases are valuable
as a source for mapping tables, which in fact can be 
complementary to \texttt{Synthesis}
for producing mapping relationships.

An issue we notice for \texttt{Synthesis}, is that while
it already distills millions of raw tables into
popular relations that requires considerable less human efforts to curate,
in some cases it still produces many somewhat redundant clusters for the same
relationship because inconsistency in value representations often lead to 
incompatible clusters that cannot be merged. Optimizing redundancy
to further reduce human efforts is a useful area for future research.

\begin{figure}[t]
\centering
\includegraphics[width=0.98\columnwidth]{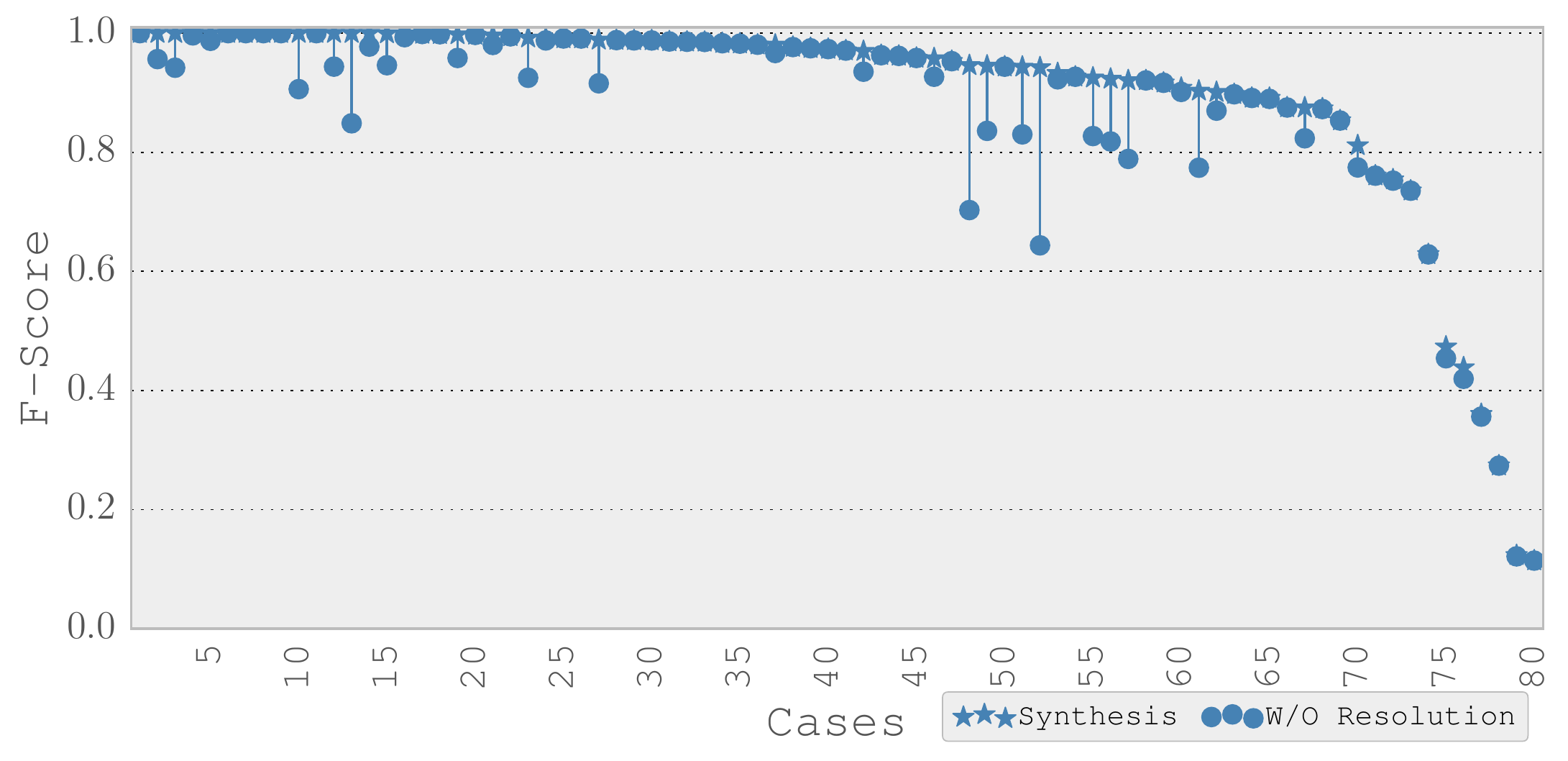}
\vspace{-3mm}
\caption{Conflict resolution improves performance.}\label{fig:resolution}
\end{figure}

%
Figure~\ref{fig:resolution} compares the f-scores with and
without conflict resolution. Conflict resolution improves the f-score in many cases.

\section{Additional Related Work}
\label{sec:additional-related}

A related problem is to discover and validate logic rules given
knowledge bases~\cite{Chen16, Galarraga13}. Our problem
is not about efficiently discovering rules that are satisfied by 
a monolithic knowledge base, instead we start from
a large set of isolated tables and we 
synthesize relationships that are functional. To some extent, the relationships
discovered by our technique can be used to by humans to 
complement and enhance existing knowledge bases.

Gupta et al. \cite{Gupta:2014} build Biperpedia, which is an 
ontology of attributes extracted from query stream and Web text. They focus on attribute name extraction for different entity classes, but not instance values in these relationships.

Mapping tables and FDs are powerful constructs that
have been studied in other contexts. For example,
authors in~\cite{Lin10, Ritter08} study the problem 
of automatically inferring functional relationships
using results extracted by Information-Extraction 
systems from a text corpus. The difficulty there is
that instances extracted for the same relation
may be inconsistent. For example, from sentences
like ``Barack Obama was born in Hawaii'' and
``Barack Obama was born in USA''
IE systems would extract ``Barack Obama''
on the left, ``Hawaii'' and ``USA'' on the
right, thus leading to the incorrect conclusion
that the relationship of birth-place is not functional.
If the results extracted by a text-pattern can be 
thought of as a table, then the task here is to
infer if FD exists for that table, and the challenge 
is that values in the table may not be consistent.
In comparison, we use tables where values are
in most cases consistent in the same column. Our
task is to go across the boundary of single tables
and produce larger relations.

While separate solutions have been proposed for certain applications 
of mapping tables such as auto-join~\cite{He15} and
auto-fill~\cite{Abedjan:2015, Yakout12}, 
we argue that there are substantial benefits for using
synthesized mapping tables. First, mapping tables are general
data assets that can benefit applications beyond auto-join and
auto-fill. Synthesizing mapping tables in essence provides a unified approach to
these related problems, instead of requiring a different solution for each
problem. Second, without mapping tables, 
techniques like~\cite{He15} perform heavy 
duty reasoning at runtime where the complexity grows quickly
for large problem instances, and thus have trouble scaling 
for latency-sensitive scenarios. In comparison, mapping table
synthesis happens offline. Applying mapping tables to online 
applications often reduces to table-lookup
that is easy to implement and efficient to scale.
Lastly, in trying to productizing auto-join and auto-fill
using techniques like~\cite{He15, Yakout12}, we notice that while
the quality are good in many cases, they can also be
unsatisfactory in others, which prevents wider adoption
in commercial systems. Synthesized mapping tables, on
the other hand, provide intermediate results that are inspectable, 
understandable, and verifiable, 
which are amenable to human curation and continuous user feedback.
Thus it is an important problem worth studying.


\end{document}